\documentclass[hyper,letterpaper,11pt]{JHEP3}
\usepackage{amsmath,amssymb,multirow,array}
\usepackage{cite}

\newcommand{\MB}{{\mathcal{M}_B}}
\newcommand{\MO}{{\mathcal{M}_\Omega}}

\newcommand{\etaH}{\tilde{\eta}}
\newcommand{\sigmaH}{\tilde{\sigma}}
\newcommand{\chiB}{\tilde{\chi}_B}
\newcommand{\chiO}{\tilde{\chi}_{\Omega}}
\newcommand{\chiE}{\tilde{\chi}_{E}}
\newcommand{\chiT}{\tilde{\chi}_{T}}

\newcommand{\zetaFI}{\beta}
\newcommand{\chiBFI}{\tilde{\beta}_B}
\newcommand{\chiOFI}{\tilde{\beta}_{\Omega}}
\newcommand{\etaFI}{\theta}
\newcommand{\etaHFI}{\tilde{\theta}}
\newcommand{\sigmaHFI}{\tilde{\kappa}_3}
\newcommand{\sigmaFI}{{\kappa}_3}
\newcommand{\chiEHFI}{\tilde{\kappa}_{2}}
\newcommand{\chiTHFI}{\tilde{\kappa}_{1}}
\newcommand{\chiEFI}{{\kappa}_{2}}
\newcommand{\chiTFI}{{\kappa}_{1}}

\newcommand{\bmu}{{\bar{\mu}}}
\newcommand{\Jsc}{J_{s\,\hbox{\tiny canon}}}
\newcommand{\cd}{\!\cdot\!}

\def\x{{\bf x}}
\def\k{{\bf k}}
\def\J{{\cal J}}
\def\T{{\cal T}}

\title{Parity-Violating Hydrodynamics in 2+1 Dimensions}

\author{Kristan Jensen$^a$, Matthias Kaminski$^{b,c}$, Pavel Kovtun$^a$, Ren\'e Meyer$^d$, Adam Ritz$^a$ and Amos Yarom$^{e}$ \\
$^a$ Department of Physics and Astronomy, University of Victoria, Victoria, BC V8W 3P6, Canada \\
$^b$ Joseph Henry Laboratories, Princeton University, Princeton, NJ 08544, USA \\
$^c$ Department of Physics, University of Washington, Seattle, WA 98195, USA \\
$^d$ Crete Center for Theoretical Physics, Department of Physics, University of Crete, 71003 Heraklion, Greece \\
$^e$ Department of Physics, Technion, Haifa 32000, Israel \\
Email: kristanj@uvic.ca, mski@uw.edu, pkovtun@uvic.ca, meyer@physics.uoc.gr, aritz@uvic.ca, ayarom@physics.technion.ac.il
}

\abstract{
We study relativistic hydrodynamics of normal fluids in two spatial dimensions. When the microscopic theory breaks parity, extra transport coefficients appear in the hydrodynamic regime, including the Hall viscosity, and the anomalous Hall conductivity. In this work we classify all the transport coefficients in first order hydrodynamics. We then use properties of response functions and the positivity of entropy production to restrict the possible coefficients in the constitutive relations. All the parity-breaking transport coefficients are dissipationless, and some of them are related to the thermodynamic response to an external magnetic field and to vorticity. In addition, we give a holographic example of a strongly interacting relativistic fluid where the parity-violating transport coefficients are computable.
}

\begin{document}

\section{Summary}
\label{S:summary}
\noindent
Hydrodynamics is an effective long-distance description of many classical or quantum many-body systems at non-zero temperature. The form of the hydrodynamic equations is dictated by the symmetries of the microscopic Hamiltonian, and is not sensitive  to the precise nature of the short-distance degrees of freedom. When the microscopic description exhibits Lorentz invariance, the collective flow is described by the relativistic analogue
of the Navier-Stokes equations. For normal fluids with an unbroken global $U(1)$ symmetry
(such as baryon number), the hydrodynamic equations take the form~\cite{LL6,weinberg:1972},
\begin{equation}
\label{eq:hydro1}
  \nabla_\mu T^{\mu\nu} = F^{\nu\mu}J_{\mu}\,,\quad\quad
  \nabla_\mu J^\mu = 0\,.
\end{equation}
Here $T^{\mu\nu}$ is the energy-momentum tensor of the fluid,  $J^\mu$
is the $U(1)$ symmetry current, and we have allowed for the possibility
of coupling the fluid to an external non-dynamical gauge field $A_{\mu}$ (with field strength $F_{\mu\nu}$) and metric $g_{\mu\nu}$ (with covariant derivative $\nabla_\mu$).
The gauge field couples to the conserved current $J^\mu$.
The relativistic analogue of the Navier-Stokes equations have a wide range of applications. For example, their study in $3+1$ dimensions has led to significant progress in
understanding the quark-gluon plasma \cite{Teaney:2009qa}.
In $2+1$ dimensions the equations were proposed as an effective description
of thermo-magnetic transport in cuprates \cite{Hartnoll:2007ih} and in graphene \cite{MS}.

A complete hydrodynamic description must, of course, supplement
equations (\ref{eq:hydro1}) with constitutive relations
which express $T^{\mu\nu}$ and $J^\mu$ in terms of  macroscopic parameters
such as the local fluid velocity $u^\mu$, local temperature $T$ and local chemical potential $\mu$. A conventional description of the constitutive relations can be found, for example, in the classic textbook by Landau and Lifshitz \cite{LL6}.
In this paper we take a closer look at the equations of relativistic hydrodynamics in $2+1$ dimensions,
and argue that the canonical constitutive relations need to be modified when the microscopic theory does not respect
parity {\bf P}. By parity we mean invariance under reflection of one of the spatial coordinates.\footnote{Parity can always be defined as reflection of one of the spatial coordinates. In a $3+1$-dimensional theory, parity combined with a rotation is equivalent to a reflection of all three coordinates. In $2+1$ dimensions, a reflection along both spatial coordinates is equivalent to a rotation.}
An example of a {\bf P}-violating system is the theory of interacting massive
Dirac fermions in 2+1 dimensions; the mass term breaks
parity.

Our expressions for the constitutive relations may be written as follows:
\begin{subequations}
\label{E:T1J1L}
\begin{eqnarray}
  &&   T^{\mu\nu} = \epsilon_0  u^\mu u^\nu
  			    +\left(P_0 - \zeta \nabla_\alpha u^\alpha - \tilde{\chi}_B B - \tilde{\chi}_\Omega \Omega\right) \Delta^{\mu\nu}
			    -\eta \sigma^{\mu\nu} - \tilde{\eta} \tilde{\sigma}^{\mu\nu} \,,  \\
  &&  J^\mu = \rho_0  u^\mu + \sigma V^{\mu} + \tilde{\sigma} \tilde{V}^{\mu} + \tilde{\chi}_E \tilde{E}^{\mu} + \tilde{\chi}_T \epsilon^{\mu\nu\rho}u_{\nu} \nabla_{\rho} T \,.
\end{eqnarray}
\end{subequations}
The tensor quantities appearing in the constitutive relations (\ref{E:T1J1L}) are
\begin{subequations}
\label{E:defs}
\begin{align}
\label{E:OandB}
	 & \Omega = -\epsilon^{\mu\nu\rho}u_{\mu} \nabla_{\nu} u_{\rho}, &
	& B = -\frac{1}{2} \epsilon^{\mu\nu\rho}u_{\mu} F_{\nu\rho}, \\
	& E^{\mu}  =  F^{\mu\nu}u_{\nu}, &
	& V^{\mu}  = E^{\mu} - T \Delta^{\mu\nu}\nabla_{\nu} \frac{\mu}{T}, \\
\label{E:sigmaDef}
	& \Delta^{\mu\nu} = u^{\mu}u^{\nu} + g^{\mu\nu}, &
	& \sigma^{\mu\nu}  = \Delta^{\mu\alpha} \Delta^{\nu\beta} \left(\nabla_{\alpha}u_{\beta} + \nabla_{\beta} u_{\alpha} - g_{\alpha\beta} \nabla_{\lambda} u^{\lambda} \right) \,,
\intertext{and}
	&\tilde{E}^{\mu}  = \epsilon^{\mu\nu\rho}u_{\nu}E_{\rho}\,,&
	&\tilde{V}^{\mu}  = \epsilon^{\mu\nu\rho}u_{\nu} V_{\rho}\,, \\
	&\tilde{\sigma}^{\mu\nu} = \frac{1}{2} \left( \epsilon^{\mu\alpha\rho} u_{\alpha} \sigma_{\rho}^{\phantom{\rho}\nu} +  \epsilon^{\nu\alpha\rho} u_{\alpha} \sigma_{\rho}^{\phantom{\rho}\mu} \right)\,.&
\end{align}
\end{subequations}
The thermodynamic parameters $P_0(\mu,T)$, $\epsilon_0(\mu,T)$ and $\rho_0(\mu,T)$ are the values of the pressure, energy density and charge density respectively in an equilibrium configuration in which $B=\Omega=0$, where $B$ is the rest-frame magnetic field and $\Omega$ the vorticity.
They satisfy
\begin{align}
\label{E:firstLawL}
dP_0 &= s_0 dT + \rho_0 d\mu\,,  \\
\label{E:eos}
\epsilon_0 &= -P_0 + s_0 T + \rho_0 \mu \,,
\end{align}
where $s_0$ is the entropy density. The velocity field is denoted $u^{\mu}$ and is normalized so that $u^{\mu}u_{\mu}=-1$. In this paper we study hydrodynamics to first order in derivatives. For counting purposes, the derivatives of $A_{\mu}$ and $g_{\mu\nu}$ are of the same order as derivatives of the hydrodynamic variables. As a result, we take the magnetic field $B$ and vorticity $\Omega$ as first order in derivatives,
and work to linear order in $B$ and $\Omega$.

The remaining parameters in \eqref{E:T1J1L} characterize the transport properties of the fluid, or its thermodynamic response. The shear viscosity $\eta$, bulk viscosity $\zeta$, and charge conductivity $\sigma$ are the canonical dissipative transport coefficients and must satisfy
\begin{equation}
\label{E:oldtransport}
	\eta \geqslant 0\,, \qquad \zeta \geqslant 0\,, \qquad \sigma \geqslant 0 \,,
\end{equation}
as a consequence of either positivity of the divergence of the entropy current, or positivity of the spectral functions in the corresponding Kubo formulas.
The Hall viscosity $\tilde\eta$ and a new parameter
$\tilde\sigma$, which contributes to the Hall effect in the absence of external magnetic fields, are both dissipationless. Our analysis does not constrain the values of $\tilde\sigma$ and $\tilde\eta$,\footnote{
When $T=0$, it has been argued that $\tilde\eta$ should coincide
with the angular momentum density of the ground state of the systems
considered in~\cite{2009PhRvB..79d5308R,Nicolis:2011ey}.
}
\begin{equation}
\label{E:newtransport}
	\tilde{\eta} \in \mathbb{R}\,,\qquad \tilde{\sigma} \in \mathbb{R} \,.
\end{equation}
The remaining parameters $\tilde{\chi}_B$, $\tilde{\chi}_{\Omega}$, $\chiE$ and $\chiT$ are not independent, and are specified in terms of three thermodynamic functions, $\MB(T,\mu)$, $\MO(T,\mu)$ and $f_\Omega(T)$, such that
\begin{subequations}
\label{E:newsusceptibilities}
\begin{align}
	\tilde{\chi}_B &= \frac{\partial P_0}{\partial \epsilon_0} \left(T\frac{\partial\MB}{\partial T}+ \mu \frac{\partial\MB}{\partial \mu} - \MB\right)
	+\frac{\partial P_0}{\partial \rho_0}\frac{\partial \MB}{\partial \mu}\,,
	\\
	\tilde{\chi}_\Omega &= \frac{\partial P_0}{\partial \epsilon_0} \left(T\frac{\partial\MO}{\partial T} + \mu \frac{\partial\MO}{\partial \mu} +f_\Omega(T) - 2\MO\right)+\frac{\partial P_0}{\partial \rho_0} \left(\frac{\partial\MO}{\partial\mu} - \MB\right)\,,
	\\
	\chiE  &= \frac{\partial \MB}{\partial \mu}-R_0\left(\frac{\partial\MO}{\partial\mu}-\MB\right) ,
	\\
	T\chiT  &= \left( T\frac{\partial\MB}{\partial T} + \mu \frac{\partial\MB}{\partial \mu} - \MB\right)
	- R_0\left(T\frac{\partial\MO}{\partial T} + \mu \frac{\partial\MO}{\partial \mu} +f_\Omega(T) -2\MO\right)\,,
\end{align}
\end{subequations}
where we have defined $R_0=\rho_0/(\epsilon_0+P_0)$. All derivatives in \eqref{E:newsusceptibilities} are evaluated at constant $\mu$ or $T$ except for $\partial P_0/\partial \epsilon_0$ and $\partial P_0/\partial\rho_0$ which are evaluated at constant $\rho_0$ and $\epsilon_0$ respectively.
The Kubo formulas for the parameters appearing in the constitutive relations \eqref{E:T1J1L} are
\begin{subequations}
\label{E:kubo0}
\begin{align}
\label{E:kubo1}
& \eta = \lim_{\omega\rightarrow 0} \frac{1}{8\omega} (\delta_{ik}\delta_{jl}-\epsilon_{ik}\epsilon_{jl}) \,{\rm Im}\, G^{ij,kl}_R(\omega,0)\,,
&  \etaH = \lim_{\omega\rightarrow 0} \frac{1}{4\omega} \delta_{ik}\epsilon_{jl}\, {\rm Im}\,G^{ij,kl}_R(\omega,0)\,,\\[5pt]
\label{E:kuboConductivity}
& \sigma = \lim_{\omega\rightarrow 0} \frac{1}{2\omega} \delta_{ij}\, {\rm Im}\,G^{i,j}_R(\omega,0)\,,
&  \sigmaH+\chiE = \lim_{\omega\rightarrow 0} \frac{1}{2\omega} \epsilon_{ij}\, {\rm Im}\, G^{i,j}_R(\omega,0)\,,
\end{align}
along with
\begin{equation}
\label{E:kubozeta}
\zeta = \lim_{\omega\rightarrow 0} \frac{1}{4\omega} \delta_{ij}\delta_{kl} \,{\rm Im}\, G^{ij,kl}_R(\omega,0)\,,
\end{equation}
\end{subequations}
and
\begin{subequations}
\label{E:kubo2}
\begin{align}
& \tilde{\chi}_B = -i \lim_{k\rightarrow 0}\frac{\epsilon_{ij}k^i}{k^2}\left(\frac{\partial P_0}{\partial \epsilon_0} G^{00,j}_R(0,k)+\frac{\partial P_0}{\partial\rho_0} G^{0,j}_R(0,k)\right), \\
&\tilde{\chi}_\Omega = -i \lim_{k\rightarrow 0}\frac{\epsilon_{ij}k^i}{k^2}\left(\frac{\partial P_0}{\partial \epsilon_0} G^{00,0j}_R(0,k)+\frac{\partial P_0}{\partial\rho_0} G^{0,0j}_R(0,k)\right),\\
& \chiE = i\lim_{k\rightarrow 0}\frac{\epsilon_{ij}k^i}{k^2}\left(G^{j,0}_R(0,k)-R_0 G^{0j,0}_R(0,k)\right), \\
& T\chiT = i\lim_{k\rightarrow 0}\frac{\epsilon_{ij}k^i}{k^2}\left(G^{j,00}_R(0,k)-R_0 G^{0j,00}_R(0,k)\right),
\end{align}
\end{subequations}
where  $\epsilon_{ij}$ is an antisymmetric tensor with $\epsilon_{12}=1$. Here $G_R(\omega,k)$ denotes the retarded Green's functions,
\begin{equation}
\nonumber
G_R^{\mu,\nu} = \langle J^{\mu}J^{\nu}\rangle_R , \qquad G_R^{\mu,\nu\rho} = \langle J^{\mu}T^{\nu\rho}\rangle_R , \qquad G_R^{\mu\nu,\rho\sigma} = \langle T^{\mu\nu}T^{\rho\sigma}\rangle_R,
\end{equation}
in the thermal equilibrium state at $B=0$ and $\Omega=0$, defined by varying the one-point functions with respect to the appropriate sources.
One important difference between the Kubo formulas \eqref{E:kubo0} and \eqref{E:kubo2} is that the former are given in terms of zero-momentum response functions, while the latter are given in terms of zero-frequency response functions.
As emphasized in \cite{Moore:2010bu}, the zero-frequency response functions are inherently Euclidean, and therefore only contain thermodynamic information.
For this reason the parameters $\chiB$, $\chiO$, $\chiE$, $\chiT$ are not transport coefficients, but should be thought of as thermodynamic quantities, consistent with \eqref{E:newsusceptibilities}.
We will refer to $\eta$, $\tilde{\eta}$, $\sigma$ and $\tilde{\sigma}$ as transport coefficients and to $\tilde{\chi}_B$, $\tilde{\chi}_{\Omega}$, $\chiE$ and $\chiT$ as thermodynamic response parameters.

Our parametrization of the constitutive relations \eqref{E:T1J1L} was not general
in the sense that we have chosen a particular out-of-equilibrium definition of energy density, charge density, and fluid velocity.
Such a choice is referred to as a ``frame''. The choice \eqref{E:T1J1L} is usually referred to as the Landau frame.
We will find it convenient to use an alternative frame that is naturally suited to fluids whose thermodynamics depends on $B$ and $\Omega$.
Indeed, a static magnetic field $B$ does not lead to an increase in (fluid) energy and therefore may be non-zero in equilibrium. Similarly, on a compact manifold, one may have non-zero vorticity $\Omega$ and still remain in thermal equilibrium, e.g., a system which executes rigid rotation.
For such equilibrium states the pressure is $P=P(T,\mu,B,\Omega)$, so that
\begin{align}
\label{E:firstLaw}
  dP& = s\, dT + \rho\, d\mu + \frac{\partial P}{\partial B}B + \frac{\partial P}{\partial \Omega} \Omega\,,\\
\label{E:eos2}
  \epsilon+P & = s T + \mu \rho\,.
\end{align}
Here and in the rest of this paper, all thermodynamic derivatives with respect to $B$ and $\Omega$ are evaluated at $B{=}0$ and $\Omega{=}0$.
The constitutive relations in a `magnetovortical' frame which is adapted to the thermodynamic relation \eqref{E:firstLaw} are given by
\begin{subequations}
\label{E:T1J1}
\begin{eqnarray}
    T^{\mu\nu} &=& \left(\epsilon - \MO\Omega + f_\Omega \Omega \right) u^\mu u^\nu \nonumber \\
 &&  			    +\left(P - \zeta \nabla_\alpha u^\alpha - \tilde{x}_B B - \tilde{x}_\Omega\, \Omega\right) \Delta^{\mu\nu}
			    -\eta \sigma^{\mu\nu} - \tilde{\eta} \tilde{\sigma}^{\mu\nu} \,,  \\
   J^\mu& =& \left(\rho - \MB\Omega \right) u^\mu + \sigma V^{\mu} + \tilde{\sigma} \tilde{V}^{\mu} + \tilde{\chi}_E \tilde{E}^{\mu} + \tilde{\chi}_T \epsilon^{\mu\nu\rho}u_{\nu} \nabla_{\rho} T \,,
\end{eqnarray}
\end{subequations}
where $\MB$ and $\MO$ are
\begin{align}
\label{E:MMO}
    \MB =  \frac{\partial P}{\partial B}\,, \qquad
	\MO =  \frac{\partial P}{\partial \Omega}
	\,.
\end{align}
The role of the undetermined function $f_{\Omega}(T)$ is unclear.
The expressions \eqref{E:newsusceptibilities} and \eqref{E:MMO}  then determine the parameters in the constitutive relations (\ref{E:T1J1}) in 
terms of thermodynamic derivatives,
\begin{subequations}
\label{E:susceptibilitiesMV}
\begin{align}
    \tilde{x}_B &= \frac{\partial P}{\partial B}\,, &
    \tilde{x}_\Omega &= \frac{\partial P}{\partial \Omega} \,,\\
    T\chiT &= \frac{\partial \epsilon}{\partial B} + R_0\left(\frac{\partial P}{\partial\Omega} - \frac{\partial \epsilon}{\partial\Omega} -f_\Omega \right)\,, &
    \chiE &= \frac{\partial \rho}{\partial B} + R_0 \left( \frac{\partial P}{\partial B} - \frac{\partial \rho}{\partial \Omega}\right)\,.
\end{align}
\end{subequations}
We note that it is also possible to present our results in a frame-invariant form along the lines of the analysis carried out in \cite{Bhattacharya:2011tr}. We describe this in Section~\ref{S:entropy}.

The constitutive relations  \eqref{E:T1J1L} (or \eqref{E:T1J1})
together with the subsequent relations for the transport coefficients and thermodynamic response parameters
are the main results of this paper. The relations \eqref{E:oldtransport} and the Kubo formulas for $\eta$, $\sigma$, and $\zeta$ are well known,
while Kubo formulas for $\tilde{\eta}$ were discussed recently in~\cite{Saremi:2011ab}. Our strategy for obtaining the relations \eqref{E:oldtransport}, \eqref{E:newtransport}, \eqref{E:newsusceptibilities}, and \eqref{E:kubo2} involved the imposition of several physical constraints on the constitutive relations. These constraints amount to requiring that the response functions of a hydrodynamic theory must: (i) obey positivity constraints, (ii) have their zero-frequency limits coincide with the corresponding thermodynamic susceptibilities, and (iii) transform covariantly under time-reversal, \textbf{T}. In addition we ensure that (iv) a local version of the second law of thermodynamics holds. Some of these constraints are more familiar than others. Relations (i) and (iv) have often been used in the literature \cite{LL6}, and (iii) is the basis for Onsager's reciprocity relations \cite{PhysRev.37.405,PhysRev.38.2265}.

Parity-violating systems in $2+1$ dimensions have been considered in the condensed matter literature. The simplest such example is  a theory of free massive Dirac fermions at zero temperature and in 2+1 dimensions.
Parity breaking leads to a
remarkable transport property at zero temperature: the Hall conductivity is quantized although no magnetic field is present~\cite{Haldane:1988zza}. This is an example of the anomalous Hall effect \cite{nsomo}. The transverse response to a thermal gradient (the thermal Hall conductivity) was recently discussed in several classes of topological insulators in 2+1 (and 3+1) dimensions, and related to anomalies in various dimensions~\cite{2010arXiv1010.0936R}.
In 2+1 dimensions the parity-odd analogue of the shear viscosity, which is called the Hall viscosity, has been studied from the condensed matter physics perspective
in~\cite{Avron:1995fg,1997physics..12050A,2009PhRvB..79d5308R,2011PhRvB..84h5316R,2011PhRvL.107g5502H},
using an effective field theory~\cite{Nicolis:2011ey,Hoyos:2011ez}, and also
using the AdS/CFT correspondence~\cite{Saremi:2011ab,Delsate:2011qp,Kimura:2011ef,Chen:2011fs}.

Parity-violating transport effects were also studied in 3+1 dimensions with QCD applications in mind: using field theory techniques~\cite{Kharzeev:2007jp,Fukushima:2008xe,Ajitanand:2010rc,Orlovsky:2010ga,Kharzeev:2010gr},
in the context of hydrodynamics~\cite{Kharzeev:2009pj,Ozonder:2010zy,KerenZur:2010zw,Kharzeev:2011ds}, and in parallel using the AdS/CFT correspondence along with hydrodynamics~\cite{Bhattacharyya:2008jc, Banerjee:2008th,Erdmenger:2008rm,Amado:2011zx,Rebhan:2009vc,Gynther:2010ed,Kalaydzhyan:2011vx,Hoyos:2011us,Matsuo:2009xn}. A relation linking parity-odd transport in 3+1 dimensions with the chiral anomaly was first found in~\cite{Son:2004tq,Son:2009tf}, and later effects of a gravitational anomaly were considered in~\cite{Landsteiner:2011iq,Landsteiner:2011cp}. Recently, parity-odd transport in superfluids was discussed in~\cite{Bhattacharya:2011tr,Neiman:2011mj,Lin:2011mr}. An effective field theory for non-dissipative transport in 1+1 dimensions was suggested in~\cite{Dubovsky:2011sk}. For considerations of hydrodynamics in arbitrary dimensions see~\cite{Kharzeev:2011ds,Loganayagam:2011mu}.

The rest of this paper is organized as follows. In Section~\ref{S:intro} we construct the most general constitutive relations allowed for a relativistic {\bf P}-violating system in 2+1 dimensions. In Section~\ref{S:entropy} we construct an entropy current with positive divergence, and determine the ensuing constraints on the constitutive relations. We independently derive restrictions on these constitutive relations in Section~\ref{S:linearized} using linearized hydrodynamics, by computing the retarded Green's functions and imposing the conditions (i)-(iii) described above. We discuss an alternative hydrodynamic frame (\ref{E:T1J1}) in Section~\ref{S:subtractions}, which provides a more transparent picture of $2+1$-dimensional thermodynamics in the presence of non-zero $B$ and $\Omega$. We check our results against an AdS/CFT computation in Section~\ref{S:holographic}, and conclude our analysis with a discussion of the results in Section~\ref{S:discussion}.

\section{The hydrodynamic expansion}
\label{S:intro}
\noindent
In hydrodynamics, the chemical potential, temperature and velocity field are allowed to vary slowly in space and time.
The four equations of motion which determine the values of the hydrodynamic variables are energy-momentum conservation and charge conservation, while the explicit relations between the energy-momentum tensor (and the current) and the hydrodynamic variables are called constitutive relations. Given a time-like vector $u^\mu$ (satisfying $u_\mu u^\mu = -1$), the energy-momentum tensor and the current
can be decomposed into pieces which are transverse and longitudinal with respect to $u$,
\begin{subequations}
\label{E:decomposition}
\begin{eqnarray}
\label{eq:T1}
  &&   T_{\mu\nu} = {\cal E} u_\mu u_\nu + {\cal P} \Delta_{\mu\nu}
       + (q_\mu u_\nu {+} q_\nu u_\mu) +t_{\mu\nu}\,,  \\
\label{eq:J1}
  &&  J_\mu = {\cal N} u_\mu + j_\mu \,,
\end{eqnarray}
\end{subequations}
where, as before, $\Delta_{\mu\nu} = g_{\mu\nu}+u_\mu u_\nu$ projects onto the space orthogonal to the velocity field.
In this decomposition, ${\cal E}$, ${\cal P}$, and ${\cal N}$ are Lorentz scalars,
$q_\mu$, $t_{\mu\nu}$, and $j_\mu$ are transverse,
$u_\mu q^\mu=0$, $u_\mu t^{\mu\nu}=0$, $u_\mu j^\mu=0$,
and $t^{\mu\nu}$ is symmetric and traceless.
Some readers may be familiar with a decomposition of the energy-momentum tensor and current into an ideal and dissipative part.
For example, one may write
\begin{align}
\begin{split}
\label{E:Constitutive2}
	T^{\mu\nu} &= \epsilon_0 u^{\mu}u^{\nu} + P_0 \Delta^{\mu\nu} + \tau^{\mu\nu}\,, \\
	J^{\mu} &= \rho_0 u^{\mu} + \Upsilon^{\mu}\,,
\end{split}
\end{align}
where $P_0$, $\epsilon_0$ and $\rho_0$ where defined below \eqref{E:defs}.
We point out that the decomposition in \eqref{E:decomposition} is of a different nature---it is a decomposition into scalar, transverse vector and transverse tensor modes which can be carried out for any tensor and vector. In other words, $\mathcal{E}$, $\mathcal{P}$ and $\mathcal{N}$ do not necessarily take on their values in equilibrium. As a consequence, the scalars, transverse vectors and transverse tensors in \eqref{eq:T1} and \eqref{eq:J1} depend, a priori, on the hydrodynamic variables $\mu$, $T$ and $u^{\mu}$ and on
quantities built from their derivatives. Needless to say, one may easily go from \eqref{E:Constitutive2} to \eqref{E:decomposition} by comparing appropriate terms in the current or energy-momentum tensor. For example: $\mathcal{N} = \rho_0 - u_{\mu}\Upsilon^{\mu}$.

Out of equilibrium one can redefine the fields $u_\mu(x)$, $T(x)$, and $\mu(x)$ in a way that simplifies the decomposition \eqref{E:decomposition}. This four-parameter field redefinition is referred to as a choice of
frame~\cite{LL6} (for a detailed recent discussion see \cite{Bhattacharya:2011ee}). In what follows we will choose a conventional Landau frame, in which the four-parameter ambiguity is fixed by requiring that $q^{\mu}=0$ and that $\mathcal{E}$ and $\mathcal{N}$ retain their values in an equilibrium configuration with zero magnetic field and zero vorticity,
i.e. $\mathcal{E}=\epsilon_0$ and $\mathcal{N}=\rho_0$.
This choice of frame gives
\begin{align}
\begin{split}
\label{eq:T2J2}
  T_{\mu\nu} &= \epsilon_0 u_\mu u_\nu + {\cal P} \Delta_{\mu\nu}
       + t_{\mu\nu}\,,  \\
  J_\mu &= \rho_0  u_\mu + j_\mu \,,
\end{split}
\end{align}
where the transverse current $j_{\mu}$ and transverse traceless tensor $t_{\mu\nu}$ vanish in the equilibrium state, and similarly
\begin{equation}
\label{E:PtoP}
	\mathcal{P} = P_0  +\left( \substack{ \hbox{\tiny first order} \\ \hbox{\tiny corrections} } \right)  \,.
\end{equation}

We now need to specify the constitutive relations which express the energy-momentum tensor and the current in terms of the hydrodynamic variables $u_\mu$, $T$, and $\mu$, their derivatives and possible electromagnetic and gravitational sources. Since we consider small deviations from thermal equilibrium, we may expand ${\cal P}$, $t_{\mu\nu}$, and $j_\mu$ to first order in derivatives of hydrodynamic variables.
As is generally the case in effective field theories, we must allow all possible one derivative contributions to ${\cal P}$, $t_{\mu\nu}$, and $j_\mu$ consistent with the symmetries of the system, but rule out those expressions that are forbidden by physical constraints such as thermodynamic laws, unitarity and time reversal symmetry.
In the remainder of this section, we will classify all possible independent contributions to $\mathcal{P}$, $j^{\mu}$, and $t^{\mu\nu}$.
By independent we mean expressions which are inequivalent under the equations of motion \eqref{eq:hydro1} to
first order in derivatives. The additional constraints which need to be implemented in order for the theory to satisfy all physical requirements will be described in Section~\ref{S:entropy} and Section~\ref{S:linearized}.

In formulating the first-order constitutive relations, we take the external fields to be small, with the field strengths
$F_{\mu\nu}$ and the connection coefficients $\Gamma_{\mu\nu}^{\rho}$
of the same order as gradients of the hydrodynamic variables.
This is the scaling required to study the
response of the fluid to sources to first order in a derivative expansion.
Magnetohydrodynamics or fluid dynamics with large values of
vorticity would require a separate treatment.

To carry out a classification of the scalars, vectors and tensors it is convenient to supplement the transverse projector $\Delta_{\mu\nu}$, satisfying $\Delta^2 = \Delta$, with a transverse antisymmetric tensor $\Sigma_{\mu\nu} \equiv \epsilon_{\mu\nu\rho}u^\rho$
which satisfies $\Sigma_{\mu\nu} u^\nu =0$, $\Sigma^2 = -\Delta$ and $\Sigma \cdot \Delta = \Sigma$. A generic vector or pseudovector can be projected into orthogonal components in the plane transverse to $u^\mu$
via  $\Delta_{\mu\nu}$ and $\Sigma_{\mu\nu}$. This allows us to straightforwardly write down all
possible structures contributing to the constitutive relations.

At first order in derivatives  there are three scalars, $u^\mu\nabla_\mu T$, $u^\mu \nabla_\mu (\mu/T)$,
 $\Delta^{\mu\nu} \nabla_\mu u_\nu = \nabla_{\mu}u^{\mu}$,
and two pseudoscalars, $\Sigma^{\mu\nu} \nabla_\mu u_\nu = -\Omega$, $\frac{1}{2} \Sigma^{\mu\nu} F_{\mu\nu} = -B$ that one could construct out of the hydrodynamic variables.
However, since there are two scalar equations of motion: $\nabla_\mu J^\mu =0$ and $u_\mu \nabla_\nu T^{\mu\nu}=u_{\mu}F^{\mu\nu}J_{\nu}$, only one of the three scalars is independent. We take this scalar to be $\Delta^{\mu\nu} \nabla_\mu u_\nu = \nabla_{\mu}u^{\mu}$. Thus, there is one scalar and two pseudoscalars which may contribute to
$\mathcal{P}$ to first order in derivatives,
\begin{subequations}
\label{E:Constitutive}
\begin{equation}
\label{eq:P1}
   {\cal P} = P_0 - \tilde{\chi}_B B
                - \tilde{\chi}_\Omega \, \Omega - \zeta\, \nabla_{\mu}u^{\mu}\,,
\end{equation}
where $P_0$ is the local thermodynamic pressure, and $\zeta$ is the bulk viscosity.
The second and third terms in \eqref{eq:P1} are forbidden in parity-invariant systems,
but are allowed once parity is broken.

Next we consider the tensors. Since $u_\mu \nabla_\nu T$ and $u_\mu \nabla_\nu (\mu/T)$ have no transverse projections,
it is sufficient to focus on projections of $\nabla_\mu u_\nu$. The two structures $\Delta^{\rho(\mu}\nabla_\rho u^{\nu)}$
and $\Sigma^{\rho(\mu}\nabla_\rho u^{\nu)}$ with circular brackets denoting a symmetric combination are, in fact, an exhaustive set of
first order tensors. Using the properties of $\Delta$
and $\Sigma$ listed above one can show that any other symmetric transverse projection can be represented as a linear combination
of these structures and $\Delta_{\mu\nu}$. Forming the trace-subtracted combinations we have, on writing out the tensors more explicitly,
\begin{align}
\begin{split}
\label{eq:pi1}
  t_{\mu\nu} =&
     - \eta\Big[\Delta_{\mu\alpha} \Delta_{\nu\beta} + \Delta_{\nu\alpha}\Delta_{\mu\beta}
               -\Delta_{\mu\nu}\Delta_{\alpha\beta}\Big] \nabla^\alpha u^\beta
      \\
     & +\frac{\etaH}{2} \Big[\Delta_{\mu\alpha} \Sigma_{\nu\beta} + \Delta_{\nu\alpha}\Sigma_{\mu\beta}
                 +\Sigma_{\mu\alpha} \Delta_{\nu\beta} + \Sigma_{\nu\alpha}\Delta_{\mu\beta}\Big]
                 \nabla^\alpha u^\beta \\
     =& - \eta\sigma_{\mu\nu} - \tilde{\eta} \tilde{\sigma}_{\mu\nu} \, ,
\end{split}
\end{align}
where $\eta$ is the shear viscosity.
The parameter $\tilde\eta$ is a {\bf P}-violating transport coefficient referred to as the Hall viscosity. It is only allowed once parity is broken, and has been discussed previously for non-relativistic~\cite{Avron:1995fg} and relativistic~\cite{Nicolis:2011ey} fluids. If we denote small fluctuations of the spatial component of the velocity field by $v^i$, then unlike the normal shear viscosity which in flat space gives a response of the stress tensor $T_{12}$ to $(\partial_1 v_2{+}\partial_2 v_1)$, the Hall viscosity
gives a response of
 $T_{12}$ to $(\partial_1 v_1{-}\partial_2 v_2)$.

There are four transverse vectors and four transverse pseudovectors which we can construct at first order in derivatives. These vectors and pseudovectors can be formed by projecting $\nabla_\mu T$, $\nabla_\mu (\mu/T)$, $F^{\mu\nu}u_{\nu} = E^{\mu}$ or $u^\nu \nabla_\nu u_\mu$ with either $\Delta_{\mu\nu}$
or $\Sigma_{\mu\nu}$. Since we have one transverse vector equation of motion, $\Delta_{\mu\nu} \nabla_\rho T^{\rho\nu}=\Delta_{\mu\nu}F^{\nu\rho}J_{\rho}$ and one transverse
pseudovector equation of motion, $\Sigma_{\mu\nu} \nabla_\rho T^{\rho\nu}=\Sigma_{\mu\nu} F^{\nu\rho} J_\rho$, only two vectors and two pseudovectors are independent.
We choose to drop the projections of $u^\nu \nabla_\nu u_\mu$.
The constitutive relation for $j^\mu$ then takes the form,
\begin{equation}
\label{eq:nu1}
  j_\mu =  \Delta_{\mu\lambda}
            \left[ \sigma V^\lambda
            +\chi_{ E} E^\lambda
            +\chi_{T} \nabla^\lambda T
            \right]
            - \Sigma_{\mu\lambda}
            \left[ \sigmaH V^\lambda
            +\tilde\chi_{ E} E^\lambda
            +\tilde\chi_{T} \nabla^\lambda T
            \right]\, ,
\end{equation}
\end{subequations}
where $V_\mu\equiv E_\mu - T \Delta_{\mu\nu}\nabla^\nu(\mu/T)$, and $E_{\mu}$ is the electric field in the fluid rest frame.

We have written the constitutive relations in the Landau frame. However, a frame-invariant definition of the transport coefficients does exist and will be discussed in the next section.

\section{Positivity of entropy production}
\label{S:entropy}
\noindent
In the Landau frame the constitutive relations take the form given in \eqref{eq:T2J2} and \eqref{E:Constitutive}.
In this section we will study how the second law of thermodynamics restricts the coefficients in the constitutive relations leading to the results \eqref{E:oldtransport}, \eqref{E:newtransport}, and \eqref{E:newsusceptibilities} described in Section~\ref{S:summary}. At intermediate stages of the computation we will find frame invariant expressions for the constitutive relations.

The assumption that the flux of entropy entering any compact spacelike region cannot exceed the amount of entropy produced in that region amounts to the existence of a current $J_{s}^{\mu}$ whose divergence is positive semi-definite,
\begin{equation}
\label{E:secondlaw}
	\nabla_{\mu} J_{s}^{\mu} \geqslant 0\,,
\end{equation}
with
\begin{equation}
	J_{s}^{\mu} = s_0 u^{\mu} + \left( \substack{\hbox{\tiny gradient} \\ \hbox{\tiny corrections} } \right)\,,
\end{equation}
where $s_0$ is the entropy density given in (\ref{E:eos}).
When there is no dissipation $J_s^{\mu}$ is conserved.
The most general form of the entropy current in a $2{+}1$-dimensional relativistic theory must take the form
\begin{equation}
	J_{s}^{\mu} = \Jsc^{\mu} + \left(\substack{ \hbox{\tiny all possible} \\ \hbox{\tiny single gradient} \\ \hbox{\tiny 3-vectors} } \right)\,,
\end{equation}
where we will refer to
\begin{equation}
\label{E:Jscanonical}
	\Jsc^{\mu}= s_0 u^{\mu} -\frac{\mu}{T} \Upsilon^{\mu} - \frac{u_{\nu}}{T}\tau^{\mu\nu}\, ,
\end{equation}
with $\tau^{\mu\nu}$ and $\Upsilon^\mu$ as in \eqref{E:Constitutive2}, as the canonical entropy current.
As we will see shortly, positivity of the divergence of the entropy current imposes
non-trivial restrictions on both $J_s^{\mu}$ and on $\tau^{\mu\nu}$ and $\Upsilon^{\mu}$ \cite{LL6}.

Our analysis closely follows \cite{Son:2009tf, Bhattacharya:2011tr}.
We have described all possible first order transverse vectors, tensors and scalars in Section~\ref{S:intro}. A list of independent transverse traceless symmetric tensors, transverse vectors, scalars and pseudoscalars is reproduced in Table~\ref{T:d1list} for convenience.
\begin{table}[tb]
\begin{center}
\renewcommand{\arraystretch}{1.5}
\begin{tabular}{| l | l | l | l | l | l |}
\hline
	scalars & pseudoscalars & transverse vectors & tensors  \\
\hline
	$ \nabla_{\mu}u^{\mu}$  &
	$\Omega = -\epsilon^{\mu\nu\rho}u_{\mu} \nabla_{\nu} u_{\rho}$ &
	$U_1^\mu =  u^{\alpha}\nabla_{\alpha} u^{\mu}  $ &
	$\sigma^{\mu\nu}$ \\
	& $B = -\frac{1}{2} \epsilon^{\mu\nu\rho}u_{\mu} F_{\nu\rho}$ &
	$U_2^\mu =   F^{\mu\nu}u_{\nu} = E^{\mu}$  & \\
	&& $U_3^{\mu}   =\Delta^{\mu\nu}\nabla_{\nu} \frac{\mu}{T} - \frac{E^{\mu}}{T} = -\frac{V^{\mu}}{T}$&\\
\hline
\end{tabular}
\caption{
\label{T:d1list}
Various independent first derivative quantities. The  shear tensor $\sigma^{\mu\nu}$ was defined in \protect\eqref{E:sigmaDef}.
Pseudotensors and pseudovectors can be obtained from the vectors and tensors above through \protect\eqref{E:pseudo}.
}
\end{center}
\end{table}
Transverse pseudovectors $\tilde{U}$ and pseudotensors $\tilde{S}$ can be obtained from the above transverse vectors and tensors via
\begin{equation}
\label{E:pseudo}
	\tilde{U}^{\mu} = \epsilon^{\mu\nu\rho}u_{\nu}U_{\rho}\,,
	\qquad
	\tilde{S}^{\mu\nu} = \frac{1}{2} \left( \epsilon^{\mu\alpha\rho} u_{\alpha} S_{\rho}^{\phantom{\rho}\nu} +  \epsilon^{\nu\alpha\rho} u_{\alpha} S_{\rho}^{\phantom{\rho}\mu} \right)\,.
\end{equation}

\begin{table}[tb]
\begin{center}
\renewcommand{\arraystretch}{1.5}
\begin{tabular}{| l |}
\hline
\multicolumn{1}{|c|}{	transverse pseudovectors}  \\
\hline
	$\tilde{V}_1^{\mu} = \epsilon^{\mu\nu\rho}u_{\nu}\nabla_\rho T = -T \tilde{U}_1^{\mu} - R_0 T^2 \tilde{U}_3^{\mu}$\\
	$\tilde{V}_2^{\mu}  = \tilde{U}_2^{\mu} $\\
	$\tilde{V}_3^{\mu}  = \epsilon^{\mu\nu\rho}u_{\nu}\nabla_{\rho} \frac{\mu}{T} = \tilde{U}_3^{\mu} + \frac{\tilde{U}_2^{\mu}}{T}$\\
	$\tilde{V}_4^{\mu}  = \frac{1}{2}\epsilon^{\mu\nu\rho} F_{\nu\rho} = \tilde{U}_2^{\mu} + u^{\mu} B$\\
	$\tilde{V}_5^{\mu}  = \epsilon^{\mu\nu\rho} \nabla_{\nu} u_{\rho} = - \tilde{U}_1^{\mu} + u^{\mu} \Omega $\\
\hline
\end{tabular}
\caption{\label{T:alternate}An alternative basis for first order pseudovectors. In relating this basis to Table~\protect\ref{T:d1list} we have used the equations of motion of ideal hydrodynamics to show that $U_1^{\mu} = -\frac{\Delta^{\mu\nu}\nabla_{\nu}T}{T} - R_0 T U_3^{\mu}$, where $R_0=\rho_0/(\epsilon_0+P_0)$.}
\end{center}
\end{table}

We will refer to the first order terms which are independent under the {equations of motion of ideal hydrodynamics} as first derivative data.
In this section we will find it convenient to use two alternative bases for the
first order pseudovectors and pseudoscalars.
The first basis is defined in \eqref{E:pseudo} and Table \ref{T:d1list}.  In addition to the basis of first-order pseudovectors spanned by $B u^\mu$,
$\Omega u^\mu$, and the $\tilde U^\mu_{i}$'s we will also find it convenient
to use a different basis of first-order pseudovectors, given in Table \ref{T:alternate}. In these bases the most general expression for the entropy current takes the form
\begin{equation}
\label{E:generalEntropyCurrent}
	J_s^{\mu} = \Jsc^{\mu} + \nu_0(\nabla \cd u) u^{\mu} + \sum_{i=1}^3 \nu_i U_i^{\mu}+ \sum_{i=1}^5 \tilde{\nu}_i \tilde{V}_i\,,
\end{equation}
where the $\nu_i$'s and the $\tilde{\nu}_i$'s are (as yet) undetermined functions of $\mu$ and $T$. By including the pseudovectors
$\tilde{V}_4$ and $\tilde{V}_5$ we have parametrized the longitudinal pseudovector contributions $u^\mu B$ and $u^\mu \Omega$, respectively, to $J_s^\mu$.
The entropy current \eqref{E:generalEntropyCurrent} contains all possible parity-even and parity-odd vector contributions both longitudinal and transverse to the velocity field.

We note in passing that
\begin{equation}
\label{E:JS-redundance}
	-\frac{\partial \tilde{\alpha}}{\partial T} \tilde{V}_1^{\mu}
	-\frac{\partial \tilde{\alpha}}{\partial \frac{\mu}{T}} \tilde{V}_3^{\mu}
	+\tilde{\alpha} \tilde{V}_5^{\mu}
	=
	 \epsilon^{\mu\nu\rho}\nabla_{\nu}\left(\tilde{\alpha} u_{\rho}\right) \, ,
\end{equation}
is a divergenceless vector for arbitrary $\tilde\alpha$. If we add such a term to the entropy current \eqref{E:generalEntropyCurrent}, it will not contribute to entropy production but it will shift $\tilde{\nu}_1$, $\tilde{\nu}_3$ and $\tilde{\nu}_5$ such that
$\tilde{\nu}_5\rightarrow \tilde{\nu}_5+\tilde{\alpha}$,
$\tilde{\nu}_1\rightarrow \tilde{\nu}_1-\partial_T \tilde{\alpha}$
and
$\tilde{\nu}_3\rightarrow \tilde{\nu}_3-\partial_{\frac{\mu}{T}}\tilde{\alpha}$. The two combinations of $\tilde{\nu}_i$'s which are invariant under this shift are  $\partial_T\tilde{\nu}_5+\tilde{\nu}_1$ and $\partial_{\frac{\mu}{T}}\tilde{\nu}_5+\tilde{\nu}_3$. Combined with $\tilde{\nu}_2$ and $\tilde{\nu}_4$, this means that there are only four combinations of the $\tilde{\nu}_i$'s that a priori can participate in entropy production.

The expression for the divergence of the entropy current in first order hydrodynamics can be written as a sum of products of first order data and a sum of genuine second order scalars,
\begin{equation}
\label{E:divergencedecomposition}
	\nabla_{\mu} J_s^{\mu} = \left(\substack{ \hbox{\tiny products of} \\ \hbox{\tiny first order data} } \right)+ \left(\substack{ \hbox{\tiny second order} \\ \hbox{\tiny scalar data} }\right) \,.
\end{equation}
By second order scalar data we mean expressions which are second order in a gradient expansion and cannot be decomposed into a product of first order terms.
Equation \eqref{E:secondlaw} states that the divergence of the entropy current should be positive semi-definite for any flow which solves the equations of motion and with any background fields. Thus, all second order data in \eqref{E:divergencedecomposition} should vanish and all first order data should arrange themselves into complete squares.

Using the scalar equation $u_{\nu}\nabla_{\mu}T^{\mu\nu} + E^\nu J_\nu + \mu\nabla_{\nu}J^{\nu}=0$,  together with \eqref{E:firstLawL} and \eqref{E:eos}, it follows that
\begin{equation}
\label{E:dmJscm}
	\nabla_{\alpha} \Jsc^{\alpha} = - \left(\nabla_{\alpha}\frac{\mu}{T} - \frac{E_{\alpha}}{T}\right) \Upsilon^{\alpha} - \nabla_{\mu} \left(\frac{u_{\nu}}{T}\right) \tau^{\mu\nu} \,.
\end{equation}
In other words, all the expressions in the divergence of the canonical part of the entropy current involve products of first order data.

All second order data in the divergence of the entropy current must vanish. Therefore the coefficients, $\nu_0$, $\nu_i$ and $\tilde{\nu}_i$ must be tuned so that no second order data appears in the divergence of the non-canonical part of $J_s^{\mu}$. Using \eqref{E:dmJscm} it is not difficult to show that
\begin{align}
\begin{split}
\label{E:entropysecond}
	\nabla_{\alpha}J_s^{\alpha} =& + \left(\nu_2 - \frac{\nu_3}{T}\right) \nabla_{\mu}E^{\mu} + \nu_3 \Delta^{\mu\nu} \nabla_{\mu}\partial_{\nu} \frac{\mu}{T} \\
	&  +	\left(\nu_0 + \nu_1\right) u^{\alpha}\nabla_{\alpha}\nabla_{\mu} u^{\mu} - \nu_1 u^{\alpha}u^{\mu}R_{\alpha\mu}\\
	& -\tilde{\nu}_2u^{\alpha}\nabla_{\alpha}B + \left(\substack{\hbox{\tiny products of} \\ \hbox{\tiny first order data}}\right)\,,
\end{split}
\end{align}
where $R_{\mu\nu}$ is the Ricci tensor. Following an analysis similar to the one carried out in \cite{Bhattacharya:2011tr} one can show that all explicit  expressions on the right hand side of \eqref{E:entropysecond} are genuine second order data. Since the genuine second order terms on the right hand side of \eqref{E:entropysecond} should vanish, we find that
\begin{equation}
\label{E:ssolini}
	\tilde{\nu}_2 = \nu_0=\nu_1  =\nu_2= \nu_3 = 0\,.
\end{equation}
There are no additional constraints that arise from demanding positivity on a curved background, so we will work in flat space from now on.

We now require that the remaining data which contributes to the divergence of the entropy current appears quadratically. We allow the undetermined variables $\tilde{\nu}_i\,,\ i=1,3,4,5$ to depend on $\bar{\mu} = \mu/T$ and $T$. To first order in the derivative expansion,
\begin{equation}
\label{E:divs1}
	\partial_{\alpha} J_s^{\alpha} = \partial_{\alpha}\Jsc^{\alpha} + \sum_{i=1,3,4,5} \left[ \frac{\partial \tilde\nu_i}{\partial T} (\partial T\cd \tilde{V}_i) + \frac{\partial \tilde\nu_i}{\partial \bar{\mu}} (\partial  \bar{\mu} \cd\tilde{V}_i) + \tilde{\nu}_i (\partial\cd\tilde{V}_i) \right]\,.
\end{equation}
Evaluating the right hand side of \eqref{E:divs1} we find
\begin{align}
\begin{split}
\label{E:divs}
	\partial_{\alpha}J_s^{\alpha} =
	&  +\partial_{\alpha} \Jsc^{\alpha} \\
	& -  \Omega(\partial \cd u) \left[ T \left(\frac{\partial P_0}{\partial\epsilon_0}\right)_{\!\rho_0} (\partial_{T}\tilde{\nu}_5+\tilde{\nu}_1) + \frac{1}{T}\left(\frac{\partial P_0}{\partial \rho_0}\right)_{\!\epsilon_0} (\partial_{\bar{\mu}}\tilde{\nu}_5+\tilde{\nu}_3)  \right] \\
	& - B(\partial \cd u) \left[ T \left(\frac{\partial P_0}{\partial\epsilon_0}\right)_{\!\rho_0}   \partial_{T} \tilde{\nu}_4 + \frac{1}{T}\left(\frac{\partial P_0}{\partial \rho_0}\right)_{\!\epsilon_0} \partial_{\bmu} \tilde{\nu}_4 \right] \\
	& +U_2 \cdot \tilde{U}_3  \left[  R_0 T (\partial_T \tilde{\nu}_3-\partial_{\bmu} \tilde{\nu}_1)  - \partial_{\bmu}\tilde{\nu}_4  + R_0 T^2 \partial_T \tilde{\nu}_4 \right] \\
	& +U_1 \cdot \tilde{U}_3 \left[ - R_0 T^2 (\partial_T\tilde{\nu}_5+\tilde{\nu}_1) + (\partial_{\bar{\mu}}\tilde{\nu}_5+\tilde{\nu}_3) + T (\partial_{\bmu} \tilde{\nu}_1   -   \partial_T \tilde{\nu}_3)  \right] \\
	& +U_1 \cdot \tilde{U}_2 \left[ \frac{\partial_{\bar{\mu}}\tilde{\nu}_5+\tilde{\nu}_3}{T} + \partial_\bmu \tilde{\nu}_1  -  \partial_T \tilde{\nu}_3 - T \partial_T \tilde{\nu}_4 \right] \,,
\end{split}
\end{align}
where we have used the identities
\begin{align}
\label{E:identities}
	u^{\alpha} \partial_{\alpha} T &= -T \left(\frac{\partial P_0}{\partial\epsilon_0}\right)_{\!\rho_0} (\partial\cd u), &
	u^{\alpha} \partial_{\alpha} \bmu &= -\frac{1}{T} \left(\frac{\partial P_0}{\partial \rho_0}\right)_{\!\epsilon_0}  (\partial\cd u),
\end{align}
derived in \cite{Bhattacharya:2011tr}. There is a redundancy in~\eqref{E:divs}: $\tilde{\nu}_1$ and $\tilde{\nu}_3$ only appear in the combinations $\partial_T\tilde{\nu}_5+\tilde{\nu}_1$ and $\partial_{\bar{\mu}}\tilde{\nu}_5+\tilde{\nu}_3$.

We now evaluate the divergence of $\Jsc^{\mu}$. Using \eqref{E:dmJscm} and \eqref{E:identities} we find
\begin{align}
\nonumber
	\partial_{\alpha} \Jsc^{\alpha}
	= & -\left(\frac{1}{2} \Delta_{\mu\nu} \tau^{\mu\nu} -  \left(\frac{\partial P_0}{\partial\epsilon_0}\right)_{\!\rho_0} u_{\mu}u_{\nu} \tau^{\mu\nu} +\left(\frac{\partial P_0}{\partial \rho_0}\right)_{\!\epsilon_0} u_{\mu} \Upsilon^{\mu} \right) \frac{\partial \cd u}{T}
	\\
\label{E:divJscSO2}
	&-\left(R_0 u_{\mu}\tau^{\mu\nu} + \Upsilon^{\nu}\right)\Delta_{\nu\alpha}U_3^{\alpha} \\
\nonumber
	& - \frac{\tau^{\mu\nu}\sigma_{\mu\nu}}{2 T}\,.
\end{align}
In this expression, $\partial{\cdot}u$ is a scalar, $\Delta_{\nu\alpha} U_3^{\alpha}$ is a transverse vector, and $\sigma_{\mu\nu}$ is a transverse traceless
symmetric tensor. Therefore, in \eqref{E:divJscSO2}, the first expression in the parentheses can be expanded in the basis of scalars, the second expression in the parentheses can be expanded in the basis of transverse vectors, and $\tau^{\mu\nu}$ can be expanded in the basis of transverse traceless symmetric tensors listed in Table~\ref{T:d1list}:
\begin{subequations}
\label{E:expansion}
\begin{align}
\label{E:expansionS}
	\frac{1}{2}\Delta_{\mu\nu} \tau^{\mu\nu} -  \left(\frac{\partial P_0}{\partial \epsilon_0}\right)_{\!\rho_0}  u_{\mu}u_{\nu} \tau^{\mu\nu} + \left(\frac{\partial P_0}{\partial \rho_0}\right)_{\!\epsilon_0} u_{\mu} \Upsilon^{\mu} =& -\zetaFI (\partial \cd u) - \chiOFI \Omega - \chiBFI B\,, \\
\label{E:expansionT}
	\Delta^{\alpha}_{\phantom{\alpha}\mu} \Delta^{\beta}_{\phantom{\beta}\nu} \tau^{\mu\nu} - \frac{1}{2} \Delta^{\alpha\beta} \Delta_{\mu\nu} \tau^{\mu\nu} =& -\etaFI \sigma^{\alpha\beta} - \etaHFI \tilde{\sigma}^{\alpha\beta}\,,
\end{align}
and
\begin{align}
\label{E:expansionV}
	\Delta^{\alpha}_{\phantom{\alpha}\nu} \left(R_0 u_{\mu} \tau^{\mu\nu} + \Upsilon^{\nu}\right) =&-T \chiTFI U_1^{\alpha}-T \chiTHFI \tilde{U}_1^{\alpha}+\chiEFI U_2^{\alpha}+\chiEHFI \tilde{U}_2^{\alpha} \\
\nonumber &- T \sigmaFI U_3^{\alpha} - T \sigmaHFI \tilde{U}_3^{\alpha}\,.
\end{align}
\end{subequations}
The expressions on the left hand side of~\eqref{E:expansion} are frame invariant. We refer the reader to  ~\cite{Bhattacharya:2011tr} for further explanation and examples. The transport coefficients on the right hand side of~\eqref{E:expansion} are physical quantities which characterize the fluid and are independent of the frame. The relations between the parameters on the right hand side of \eqref{E:expansion} and the coefficients in the Landau frame are given in \eqref{E:ToLandau}.

Requiring that the divergence of the entropy current be positive semi-definite leads to the following constraints:
\begin{align}\nonumber
	\zetaFI &\geqslant 0\,, \\
\label{E:resultS}
	\chiOFI &=\left(\frac{\partial P_0}{\partial\epsilon_0}\right)_{\rho_0} T^2(\partial_T\tilde{\nu}_5+\tilde{\nu}_1)+\left(\frac{\partial P_0}{\partial \rho_0}\right)_{\epsilon_0}( \partial_{\bar{\mu}}\tilde{\nu}_5+\tilde{\nu}_3 )\,, \\
\nonumber
	\chiBFI & = \left(\frac{\partial P_0}{\partial\epsilon_0}\right)_{\rho_0} T^2\partial_T\tilde{\nu}_4+\left(\frac{\partial P_0}{\partial \rho_0}\right)_{\epsilon_0} \partial_{\bar{\mu}}\tilde{\nu}_4 \,,
\end{align}
in the scalar sector,
\begin{subequations}
\label{E:resultV}
\begin{align}
	\chiTFI=0\,,\qquad	\chiEFI =0\,,\qquad	 \sigmaFI \geqslant 0\,,\qquad \sigmaHFI \in \mathbb{R}\,,
\end{align}
as well as
\begin{align}
	\chiTHFI&= T(\partial_T\tilde{\nu}_4 -R_0(\partial_T\tilde{\nu}_5+\tilde{\nu}_1))\,, &\chiEHFI&=\partial_{\bar{\mu}}\tilde{\nu}_4-R_0(\partial_{\bar{\mu}}\tilde{\nu}_5+\tilde{\nu}_3)\,,
\end{align}
\end{subequations}
in the vector sector, and
\begin{align}
\begin{split}
\label{E:resultT}
	\etaFI \geqslant 0\,, \qquad	\etaHFI \in \mathbb{R} \, ,
\end{split}
\end{align}
in the tensor sector. A further condition is
\begin{equation}
\label{E:Constraint} \partial_{\bar{\mu}}\tilde{\nu}_5+\tilde{\nu}_3 + T(\partial_{\bmu} \tilde{\nu}_1 - \partial_T \tilde{\nu}_3)  - T^2 \partial_T \tilde{\nu}_4 = 0\,,
\end{equation}
which we used to simplify the expressions for $\chiTHFI$ and $\chiEHFI$ in~\eqref{E:resultV}.

In the discussion below equation \eqref{E:JS-redundance} we argued that only four combinations of $\tilde{\nu}_i$ participate in entropy production. Since we found that $\tilde{\nu}_2=0$, three combinations remain: $(\partial_T \tilde{\nu}_5 + \tilde{\nu}_1)$, $(\partial_{\bar\mu} \tilde{\nu}_5 + \tilde{\nu}_3)$, and $\tilde{\nu}_4$. Equation \eqref{E:Constraint} may be rewritten in terms of these three combinations,
\begin{equation}
\label{E:ConstraintPhys}
\left(\partial_{\bar{\mu}}\tilde{\nu}_5 + \tilde{\nu}_3 \right)+\frac{1}{T} \partial_{\bar{\mu}} \left(T^2 \left(\partial_T \tilde{\nu}_5 + \tilde{\nu}_1 \right)\right) -T\partial_T \left(\partial_{\bar{\mu}}\tilde{\nu}_5 + \tilde{\nu}_3 \right) - T^2 \partial_T \tilde{\nu}_4 = 0\, .
\end{equation}
We now parametrize the coefficients $\tilde{\nu}_4$ and $(\partial_{\bar{\mu}}\tilde{\nu}_5 +\tilde{\nu}_3)$ as
\begin{equation}
\label{E:nuToM}
T\tilde{\nu}_4 = \MB, \qquad \partial_{\bar{\mu}}\tilde{\nu}_5 +\tilde{\nu}_3=\frac{1}{T}\partial_{\bar{\mu}}\MO-\MB,
\end{equation}
where $\MB$ and $\MO$ are arbitrary functions of $\mu$ and $T$. Relation~\eqref{E:ConstraintPhys} then takes the form
\begin{equation}
\frac{1}{T}\frac{\partial}{\partial\bar{\mu}}\left(T^2(\partial_T\tilde{\nu}_5+\tilde{\nu}_1) -T\partial_T\MO+2\MO\right)=0\,,
\end{equation}
which has the solution
\begin{equation}
\label{E:diffrelation}
T^2(\partial_T\tilde{\nu}_5+\tilde{\nu}_1) = T\partial_T \MO-2\MO+f_\Omega(T)\, ,
\end{equation}
where $f_\Omega(T)$ is undetermined.

The Landau frame was defined by \eqref{eq:T2J2}, \eqref{eq:P1}, \eqref{eq:pi1} and \eqref{eq:nu1}. Inserting the Landau-frame expressions for the stress tensor and charge current into \eqref{E:expansion} we find
\begin{subequations}
\label{E:ToLandau}
\begin{align}
    \etaFI &= \eta \, , &
    \etaHFI &= \tilde{\eta}\, , &
    \zetaFI &= \zeta \, , &  \\
    \chiTFI &= \chi_T\, ,&
    \chiEFI &= \chi_E \,,&
    \sigmaFI &= \sigma+R_0 T \chi_T \, , \\
    \chiTHFI &= \chiT\, ,&
    \chiEHFI &= \chiE\, ,&
    \sigmaHFI &= \tilde{\sigma}+R_0 T \chiT  \,, \\
    \chiBFI &= \chiB\, ,&
    \chiOFI &= \chiO\, . &&
\end{align}
\end{subequations}
Inserting \eqref{E:nuToM} and \eqref{E:diffrelation} into~\eqref{E:resultS} and~\eqref{E:resultV} and converting to the Landau frame \eqref{E:ToLandau} yields the expressions~\eqref{E:newsusceptibilities} for the $\tilde{\chi}$'s as functions of $\MB(T,\mu)$, $\MO(T,\mu)$ and $f_\Omega(T)$.

\section{Response functions}
\label{S:linearized}
The theory of linear response allows one to relate transport properties and thermodynamic susceptibilities to limiting values of retarded correlation functions~\cite{JPSJ.12.570,KM}. In Section \ref{S:GreenGeneral} we will discuss several properties of the retarded functions for fluids.
These include positivity of spectral functions, covariance under time reversal, and connections to thermodynamic susceptibilities in appropriate limits.  Then in Sections \ref{S:PEven} and \ref{S:Podd}, we will show how these properties may be used to restrict the coefficients which appear in the constitutive relations.

\subsection{General properties}
\label{S:GreenGeneral}

Consider the expressions for the current and the energy-momentum tensor densities in the presence of external sources,
\begin{align}
&  \J^\mu(x) \equiv \sqrt{-g}\; \langle J^{\mu}(x)\rangle_{A,g}\,,
&  \T^{\mu\nu}(x) \equiv \sqrt{-g}\; \langle T^{\mu\nu}(x)\rangle_{A,g}\,.
\label{eq:JTag}
\end{align}
where $x=(t,\x)$, $g\equiv{\rm det}(g)$ and $g_{\mu\nu}$ is the metric.
The subscripts $A$ and $g$ indicate that these are one-point functions
in the presence of a background gauge field $A_{\mu}$
and a metric $g_{\mu\nu}=\eta_{\mu\nu} + h_{\mu\nu}$,
with $\eta_{\mu\nu}$ the mostly-plus flat-space metric.
When the background sources vary sufficiently slowly in space and time,
the assumption of the hydrodynamic theory is that $\langle J^{\mu}(x)\rangle$ and $\langle T^{\mu\nu}(x)\rangle$ in~\eqref{eq:JTag} are given precisely by the constitutive relations (\ref{eq:T2J2}), (\ref{E:Constitutive}),
with the hydrodynamic variables $\epsilon_0$, $u^\mu$, and $\rho_0$
satisfying the conservation equations (\ref{eq:hydro1}).
To obtain retarded two-point functions in flat space and without external fields,
we need to solve (\ref{eq:hydro1}) to first order in $A_\mu$ and $h_{\mu\nu}$, insert those solutions into the constitutive relations,
and then differentiate (\ref{eq:JTag}) with respect to the sources.
We define the retarded functions:%
\footnote{%
		More formally, one can start with the generating functional
		in the Schwinger-Keldysh formalism \cite{Chou:1984es,Wang:1998wg}
		and define the fully retarded functions by taking the
		appropriate variations as is done in, for example, \cite{Moore:2010bu}.
}
\begin{subequations}
\label{E:GreenDef}
\begin{align}
\label{E:GRmnDefW}
	G_R^{\mu,\nu}(x) & =
	\left.\frac{\delta \J^{\mu}(x) }{\delta A_{\nu}(0)}\right|_{A=h=0}\,, &
	G_R^{\mu\nu,\sigma}(x) & =
	\left.\frac{\delta \T^{\mu\nu}(x)}{\delta A_{\sigma}(0)}\right|_{A=h=0}\,, \\[5pt]
	G_R^{\sigma,\mu\nu}(x) & =2
	\left.\frac{\delta \J^{\sigma}(x) }{\delta h_{\mu\nu}(0)}\right|_{A=h=0}\,, &
	G_R^{\sigma\tau,\mu\nu}(x) & = 2
	\left.\frac{\delta \T^{\sigma\tau}(x) }{\delta h_{\mu\nu}(0)}\right|_{A=h=0}\,.
\end{align}
\end{subequations}
Since we will be interested in the retarded functions in the equilibrium state
which has $B=0$ and $\Omega=0$,
we need only solve the hydrodynamic equations of motion to first order in
fluctuations in order to find $J_\mu$ and $T_{\mu\nu}$ to first order in background fields.

Alternatively, the retarded functions may be defined in the canonical formalism,
\begin{subequations}
\label{E:GRmnDefC}
\begin{align}
	G_R^{\mu,\nu}(x) & = i \theta(t)\,
	\hbox{Tr}\left( \varrho \left[J^{\mu}(x),\,J^{\nu}(0)\right] \right), &
	G_R^{\mu\nu,\sigma}(x) & = i \theta(t)\,
	\hbox{Tr}\left( \varrho \left[T^{\mu\nu}(x),\,J^{\sigma}(0)\right] \right)\,,\\[5pt]
	G_R^{\sigma,\mu\nu}(x) & = i \theta(t)\,
	\hbox{Tr}\left( \varrho \left[J^{\sigma}(x),\,T^{\mu\nu}(0)\right] \right), &
	G_R^{\sigma\tau,\mu\nu}(x) & = i \theta(t)\,
	\hbox{Tr}\left( \varrho \left[T^{\sigma\tau}(x),\,T^{\mu\nu}(0)\right] \right)\,,
\end{align}
\end{subequations}
where $\varrho$ is the grand canonical density operator
specifying the equilibrium state of the system. The retarded functions defined by \eqref{E:GreenDef} and \eqref{E:GRmnDefC} will differ by contact terms. See for example \cite{Policastro:2002tn,Romatschke:2009ng} for a discussion.

In the remainder of this subsection we will study general properties of the retarded functions defined in~\eqref{E:GreenDef}. In the zero-frequency limit, the retarded functions may be determined by equilibrium thermodynamics. Indeed, the retarded functions at zero frequency coincide with the Euclidean time-ordered functions at zero frequency, and can be computed for example by the Euclidean functional integral method in the grand canonical ensemble.

In a static equilibrium without external sources, the partition function $Z$ in the grand canonical ensemble is given by
\begin{equation}
\label{E:Zdef}
    Z[T,\,\mu] = \hbox{Tr}\left[\exp
        \left(-\frac{H}{T} + \frac{\mu Q}{T}  \right)
        \right]\,,
\end{equation}
where $H$ is the Hamiltonian of the system and $Q$ its total charge. In this coordinate frame the fluid velocity is
 $u^{i}=0$.
When external sources are turned on, an interesting feature of the equilibrium partition function is that constant sources $A_0$, $h_{00}$ and $h_{0i}$ may be eliminated by a suitable redefinition of thermodynamic variables. To see this, consider the Euclidean theory. Let $\tau$ be the Euclidean time with period $\beta = 1/T$. The chemical potential may be defined through the Wilson loop around the time circle
\begin{equation}
\label{E:wilsonMu}
\frac{\mu}{T} = i\int_0^{\beta}d\tau \,A_{\tau}\,.
\end{equation}
Suppose that the fluid is subjected to constant sources $A_0$, $h_{00}$, and $h_{0i}$.
A constant contribution to $A_0$ shifts the chemical potential such that $\mu'/T' = \mu/T+A_0/T$.
A constant $h_{00}$ can be eliminated in a similar manner by rescaling $\tau$. Since the length of the time circle does not change under a rescaling, the periodicity of $\tau$ must be shifted to $\beta' = \beta(1-h_{00}/2)$ to first order in $h_{00}$.
The temperature and chemical potential are then
\begin{equation}
T' = T\left(1+\frac{h_{00}}{2}\right)\,, \qquad \mu' = \mu\left(1+\frac{h_{00}}{2}\right)+A_0\,,
\end{equation}
to first order in external fields. Finally, a constant perturbation of the Minkowski metric $h_{0i}$ may be removed by a coordinate transformation. The fluid velocity $u^i$ is invariant under this transformation, to first order in $h_{0i}$. Thus, the partition function in the presence of constant $A_0$, $h_{00}$, and $h_{0i}$ takes the form
\begin{equation}
\label{E:ZSources}
     Z[T,\,\mu ;\,A_0,\,h_{00},\,h_{0i}] = Z\left[T\left(1+\frac{h_{00}}{2} \right),\,\mu\left(1+\frac{h_{00}}{2}\right)+A_0 ;\,0,\,0,\,0\right]\,,
\end{equation}
and the fluid velocity $u^i$ is unchanged.
Put differently, the thermodynamic behavior of a system with constant and small $A_0$, $h_{00}$ and $h_{0i}$ is equivalent to the thermodynamic behavior of the same system with zero $A_0$, $h_{00}$ and $h_{0i}$ but appropriately shifted temperature, chemical potential, and normalized three-velocity field.
Retarded functions at zero frequency may then be evaluated in static equilibrium with $u^i=0$. By using the expression for the generating function on the right hand side of \eqref{E:ZSources} as well as the constitutive relations~\eqref{E:Constitutive2} we find
\begin{subequations}
\label{eq:susceptibility-constraints-1}
\begin{align}
\label{suscept1}
	 \lim_{\k\to0}G_R^{0,0}(\omega{=}0,\k)  &=
	  \left(\frac{\partial\rho_0}{\partial \mu}\right)_{\!\!T},
	& \lim_{\k\to0}G_R^{0,00}(\omega{=}0,\k)  &=
	 T\!\left(\frac{\partial\rho_0}{\partial T}\right)_{\!\!\mu/T} \,,\\[5pt]
\label{suscept2}
	 \lim_{\k\to0}G_R^{00,0}(\omega{=}0,\k)  &=
	 \left(\frac{\partial\epsilon_0}{\partial \mu}\right)_{\!\!T},
	& \lim_{\k\to0}G_R^{00,00}(\omega{=}0,\k)  &=
	  T\!\left(\frac{\partial\epsilon_0}{\partial T}\right)_{\!\!\mu/T}\,.
\end{align}
\end{subequations}
We will refer to the relations~\eqref{eq:susceptibility-constraints-1}
as susceptibility conditions, and impose them as constraints that the hydrodynamic retarded functions evaluated later in this section must satisfy.

The response functions defined in~\eqref{E:GRmnDefC} are also constrained by their behavior under time-reversal {\bf T} which is the basis for the Onsager relations \cite{PhysRev.37.405,PhysRev.38.2265}.
For two local Hermitian operators ${\cal O}_1$ and ${\cal O}_2$
which transform in a definite way under time reversal,
$\Theta {\cal O}_{i} \Theta^{-1} = n_{i} {\cal O}_{i}$,
the anti-unitarity of the time-reversal operator $\Theta$
combined with translation invariance implies
\begin{align}
\label{TcovRS}
    G_R^{ij}(x) \equiv i\theta(t)\,
	\hbox{Tr} \left( \varrho [\mathcal{O}_i(t,\x),\mathcal{O}_j(0)] \right)
	= i\theta(t)\,
	n_i n_j \hbox{Tr} \left( \varrho' [\mathcal{O}_j(t,-\x),\,\mathcal{O}_i(0)] \right),
\end{align}
where $\varrho'=\Theta\varrho\Theta^{-1}$ is the time reversed density operator.
We allow for the possibility that {\bf T} may be broken
in the microscopic theory
by a set of real parameters such as fermion masses or the magnetic field,
which we collectively denote as $b_a$. The transformation ${\bf T}'$, which is ${\bf T}$ combined with $b_a\to-b_a$, is then a symmetry, so that $\varrho'(b_a) = \varrho(-b_a)$.
The Fourier transform of the retarded function must then satisfy
\begin{equation}
\label{TcovFS}
 G_{R}^{ij}(\omega,\k; b_a) = n_i n_j G_{R}^{ji}(\omega,-\k; -b_a)\,,
\end{equation}
which will impose constraints on the transport coefficients. 

The canonical definition of the retarded function implies
\begin{equation}
\label{unitCon}
\text{Im}\,G_{R}^{ii}(\omega,\k) \geqslant 0\,,
\end{equation}
for $\omega\geqslant0$.
When applied to the Kubo formulas for transport coefficients,
this condition can be used to argue that the viscosities $\eta$ and $\zeta$,
and the conductivity $\sigma$ cannot be negative. This agrees with the argument \cite{LL6}
that $\eta$, $\zeta$, and $\sigma$ contribute to entropy production and therefore cannot be negative.
However, \eqref{unitCon} does not constrain
the sign of their parity-odd cousins $\etaH$, $\chiO$, and $\sigmaH$.

\subsection{Parity-preserving hydrodynamics}
\label{S:PEven}
We first discuss the simpler case of fluids where parity is a symmetry of the microscopic theory. The  constitutive relations in the Landau frame are given by
\eqref{eq:T2J2} and \eqref{E:Constitutive}
with the parity-odd terms omitted,
\begin{align}
	T^{\mu\nu} &= \epsilon_0 u^{\mu}u^{\nu} + P_0 \Delta^{\mu\nu} - \zeta \nabla_{\mu}u^{\mu} - \eta \sigma^{\mu\nu}\,, \\
	J^{\mu} & = \rho_0 u^{\mu} + \Delta^{\mu\lambda}
            \left[ \sigma V_\lambda
            +\chi_{ E} E_\lambda
            +\chi_{T} \nabla_\lambda T
            \right]\,,
\end{align}
where $E^\mu = F^{\mu\nu}u_\nu$,
$V_\mu = E_\mu - T \Delta_{\mu\nu}\nabla^\nu(\mu/T)$ and
\begin{equation}
	\sigma_{\mu\nu} = \Big[\Delta_{\mu\alpha} \Delta_{\nu\beta} + \Delta_{\nu\alpha}\Delta_{\mu\beta} -\Delta_{\mu\nu}\Delta_{\alpha\beta}\Big] \nabla^\alpha u^\beta \,.
\end{equation}
While the constitutive relations are written in a particular frame, the correlation functions we obtain are frame invariant. They depend on the physical parameters defined in~\eqref{E:expansion}. In the parity-preserving case and in the Landau frame, these physical parameters are $\eta,\zeta,\sigma$, $\chi_E$, and $\chi_T$.

As we have shown in Section \ref{S:entropy}, the positivity of entropy produciton
implies that $\eta$, $\zeta$, and $\sigma$ are non-negative, while $\chi_T$ and $\chi_E$ both vanish. Alternatively, the same results follow from the properties of the retarded two-point functions described in Section~\ref{S:GreenGeneral}. We will see that the conditions
\begin{equation}
\label{E:Restrictions2}
	\eta\geqslant0\,, \quad
	\zeta\geqslant0\,, \quad
	\sigma\geqslant0\,,
\end{equation}
follow from the positivity of the spectral function \eqref{unitCon},
while the conditions
\begin{equation}
\label{E:Restrictions1}
	\chi_T=0\,, \quad
	\chi_E=0,
\end{equation}
follow from the susceptibility constraints \eqref{eq:susceptibility-constraints-1}.

To compute the retarded functions we need to solve the linearized hydrodynamic equations.
The equilibrium state is described by a stationary and homogenous solution
to the equations of motion \eqref{eq:hydro1}
which has constant energy density $\epsilon_0$, constant pressure $P_0$,
the velocity field $u^{\mu}=(1,0,0)$,
and vanishing sources.
We solve the linearized equations of motion \eqref{eq:hydro1} in the presence of external sources $h_{\mu\nu}$ and $A_\mu$ for $\delta \mu$, $\delta T$, $\delta u^1$ and $\delta u^2$ which specify the linearized corrections  to the chemical potential, temperature, and spatial components of the velocity field respectively.

The linearized hydrodynamic equations of motion may be written as
\begin{align}
\label{PevenHydro}
&D X = \mathcal{S} \,,
\end{align}
where $X = (\delta\mu,\delta T, \delta u^1, \delta u^2)$
is a vector of linearized hydrodynamic variables,
$D$ is a matrix of second-order differential operators,
and $\mathcal{S}$ is a vector built out of the external sources
$h_{\mu\nu}$, $A_\mu$, and their derivatives.
This system of differential equations may be Fourier transformed
to obtain a set of algebraic equations for the Fourier components
of the hydrodynamic variables.
Taking the spatial momentum $\k$ in the $x^1$ direction gives
\begin{equation}\hspace{-0.5cm}
D = \left(\begin{array}{cccc}
k^2\sigma - i \omega \frac{\partial\rho_0}{\partial \mu} & -k^2\left( \frac{\mu}{T}\sigma+\chi_T\right) - i \omega \frac{\partial\rho_0}{\partial T} & i k \rho_0 & 0 \\[5pt]
-i\omega \frac{\partial \epsilon_0}{\partial \mu} & -i \omega\frac{\partial\epsilon_0}{\partial T} & i k (\epsilon_0{+}P_0) & 0 \\[5pt]
i k \rho_0 & i k s_0 & k^2(\eta{+}\zeta)-i\omega(\epsilon_0{+}P_0) & 0 \\[5pt]
0 & 0 & 0 & k^2\eta - i \omega(\epsilon_0{+}P_0)
\end{array}\right),
\end{equation}
where $s_0$ is the entropy density, $k\equiv |\k|$,
and all derivatives are evaluated at constant $\mu$ or $T$.
The eigenvectors of the matrix $D$ correspond to the hydrodynamic modes of the theory,
whose dispersion relations may be obtained by solving for the roots of the determinant of $D$.

Before presenting explicit expressions for the correlation functions,
we pause to note that we are working with hydrodynamics to first order in derivatives,
ignoring possible two-derivative terms in the constitutive relations.%
\footnote{
	Second-order hydrodynamics by itself is not a consistent effective description
	of fluids, as the infrared effects of hydrodynamic fluctuations turn out to be more important
	than those due to two-derivative terms in the constitutive relations.
	See for example Ref.~\cite{Kovtun:2011np} for a discussion.
	We will ignore these fluctuation effects in this paper,
	which may be consistently done in the large-$N$ limit~\cite{Kovtun:2003vj}.
}
This means that equations \eqref{PevenHydro} are valid only to second order
in gradients of the hydrodynamic variables.
In linearized hydrodynamics, possible two-derivative terms in the constitutive relations
may be accounted for if we replace
$\mathcal{S}$ with $\mathcal{S} + \mathcal{O}(\partial^3)$, and
$D$ with $D + \mathcal{O}(\partial^3)$ where by $\mathcal{O}(\partial^3)$
we mean expressions whose combined powers of $k$ and $\omega$
are greater than or equal to three.
For example, the fluctuation of the transverse velocity $\delta u^2$ takes the form
\begin{equation}
	\delta u^2[\mathcal{S}] = \frac{\mathcal{S}_4 + \mathcal{O}(\partial^3)}
	                  {k^2 \eta - i\omega(\epsilon_0{+}P_0)+\mathcal{O}(\partial^3)}\,.
\end{equation}
We use parameterizations of this form to ensure that our results are free from these second-order corrections.

To impose the susceptibility condition \eqref{suscept1} we compute the density-density function $G_R^{0,0}(\omega,k)$ which can be obtained by solving
the equations of motion~\eqref{PevenHydro} and using them to evaluate the one-point function $\J^0[\mathcal{S}]$.
Differentiating the resulting $\J^0[\mathcal{S}]$ with respect to $A_0$
according to the definition~\eqref{E:GRmnDefW}, we find
\begin{multline}
\label{G00}
  G_R^{0,0} (\omega,k) =\frac{k^2 (k^2\eta - i \omega (\epsilon_0{+}P_0)) }{T\det(D)}
  \Bigg\{ k^2(\epsilon_0{+}P_0) \Bigg[\sigma(\epsilon_0{+}P_0)
  +\chi_E T \left( s_0  \frac{\partial\rho_0}{\partial\mu} -
  \rho_0\frac{\partial s_0}{\partial\mu}\right) \\
  +\chi_T T\rho_0\frac{\partial\rho_0}{\partial\mu} \Bigg]
  - i\omega |\mathfrak{X}|T^2 \left({\rho_0}^2+(\sigma{+}\chi_E)(k^2(\eta{+}\zeta)
  + i\omega (\epsilon_0{+}P_0))\right)\Bigg\},
\end{multline}
where
\begin{equation}
	|\mathfrak{X}| = \frac{\partial \rho_0}{\partial \mu} \frac{\partial s_0}{\partial T} - \frac{\partial\rho_0}{\partial T} \frac{\partial s_0}{\partial \mu}
\end{equation}
is the determinant of the thermodynamic susceptibility matrix.
One can explicitly check that the $\mathcal{O}(\partial^3)$ terms
do not contribute to the zero frequency, small momentum, limit of $G_R^{0,0}(\omega,k)$
and therefore we can reliably use \eqref{G00} to compute
\begin{equation}
\label{E:G00k}
\lim_{k\rightarrow 0}G_R^{0,0}(\omega{=}0,k) = \frac{\partial\rho_0}{\partial\mu}+\frac{T\left(s_0 \frac{\partial \rho_0}{\partial \mu} - \rho_0 \frac{\partial s_0}{\partial \mu}\right)}{(\epsilon_0{+}P_0)\sigma+T\rho_0\chi_T}\chi_E\,.
\end{equation}
The susceptibility constraint \eqref{suscept1} then implies
$
	\chi_E=0\,.
$
Similarly, a straightforward computation gives
\begin{equation}\hspace{-0.1cm}
\label{E:G000k}
\lim_{k\rightarrow 0}G_R^{0,00}(\omega{=}0,k) = \frac{(\epsilon_0+P_0) \sigma}{(\epsilon_0+P_0)\sigma + T \rho_0 \chi_T} T \left(\frac{\partial \rho_0}{\partial T}\right)_{\frac{\mu}{T}} + \frac{(\epsilon_0+P_0) T \chi_T}{(\epsilon_0+P_0)\sigma + T \rho_0 \chi_T} \left(\frac{\partial \rho_0}{\partial \mu}\right)_{T}\,.
\end{equation}
Using the second equation in \eqref{suscept1}, one finds $\chi_T=0$.

For the retarded function $G_R^{12,12}$ one finds
\begin{equation}
G_R^{12,12}(\omega,k{=}0) = -P+i\eta\omega + \mathcal{O}(\omega^2)\,.
\end{equation}
The positivity condition \eqref{unitCon} applied to the spectral function of $T^{12}$ then implies that $\eta\geqslant 0$. An identical analysis with the spectral functions of $T^{11}$ and $j^1$ shows that $\zeta\geqslant 0$ and $\sigma\geqslant0$. We have not found any other restrictions which follow from our consistency conditions.

\subsection{Parity-violating hydrodynamics}
\label{S:Podd}
We now move on to parity-violating hydrodynamics in 2+1 dimensions. Equations~\eqref{eq:T2J2} and~\eqref{E:Constitutive} give the most general constitutive relations in the Landau frame in the absence of parity as a symmetry.
The linearized equations take the form \eqref{PevenHydro} with
\begin{equation}\hspace{-0.5cm}
D = \left(\begin{array}{cccc}
k^2\sigma - i \omega \frac{\partial\rho_0}{\partial \mu} & -k^2\left( \frac{\mu}{T}\sigma+\chi_T\right) - i \omega \frac{\partial\rho_0}{\partial T} & i k \rho_0 & 0\\[5pt]
-i\omega \frac{\partial \epsilon_0}{\partial \mu} & -i \omega\frac{\partial\epsilon_0}{\partial T} & i k (\epsilon_0{+}P_0) &0 \\[5pt]
i k \rho_0 & i k s_0 & k^2(\eta{+}\zeta)-i\omega(\epsilon_0{+}P_0) & k^2\left(\chiO{+}\etaH\right) \\[5pt]
0 & 0 & -k^2\etaH & k^2\eta - i \omega(\epsilon_0{+}P_0)
\end{array}\right)\,.
\end{equation}
Imposing the susceptibility constraints \eqref{suscept1} implies, via a computation identical to \eqref{E:G00k} and \eqref{E:G000k}, that $\chi_E=0$ and $\chi_T=0$, exactly as in the parity-preserving hydrodynamics.

To obtain Kubo formulas for the $\tilde{\chi}$'s, it is useful to consider the combinations
\begin{equation}
\mathcal{C}^0 = \left(\frac{\partial P_0}{\partial\rho_0}\right)_{\!\epsilon_0} \mathcal{J}^0+\left(\frac{\partial P_0}{\partial\epsilon_0}\right)_{\!\rho_0} \mathcal{T}^{00}, \qquad \mathcal{C}^i = \mathcal{J}^i - R_0 \mathcal{T}^{0i},
\end{equation}
where $R_0=\rho_0/(\epsilon_0{+}P_0)$.
A direct computation shows that at zero frequency, the $\mathcal{O}(k)$ parts of $\mathcal{C}^{0}$, $\mathcal{C}^{i}$ and $T^{ii}$ are reliably evaluated in first-order hydrodynamics.
We then find the following relations for the retarded functions at $\omega{=}0$:
\begin{subequations}
\label{E:kuboTildeChi}
\begin{align}
\label{E:kuboTB}
\lim_{k\rightarrow 0}\frac{1}{ik}\langle\mathcal{C}^0 \mathcal{J}^2\rangle_R(0,k) &= \chiB\,, &\lim_{k\rightarrow 0}\frac{1}{ik}\langle\mathcal{C}^0 \mathcal{T}^{02}\rangle_R(0,k) & = \chiO\,, \\
\label{E:kuboTE}
\lim_{k\rightarrow 0}\frac{1}{ik}\langle\mathcal{C}^2 \mathcal{J}^0\rangle_R(0,k) & = -\chiE\,, &\lim_{k\rightarrow 0}\frac{1}{ik}\langle \mathcal{C}^2 \mathcal{T}^{00}\rangle_R(0,k) & = - T \chiT, \\
\label{E:tii}
\lim_{k\rightarrow 0}\frac{1}{ik}\langle \mathcal{T}^{ii}\mathcal{J}^2\rangle_R(0,k) & =0\,, & \lim_{k\rightarrow 0}\frac{1}{k}\langle \mathcal{T}^{ii}\mathcal{T}^{02}\rangle_R(0,k) & =0\,.
\end{align}
\end{subequations}
Note that in \eqref{E:tii} no summation over $i$ is implied. For general $\mathbf{k}$ equations~\eqref{E:kuboTB} and \eqref{E:kuboTE} become the Kubo formulas~\eqref{E:kubo2} written in Section~\ref{S:summary}. The last line will be relevant later in Section~\ref{S:subtractions}.

The $\tilde{\chi}$'s are also related by an Onsager relation. Using~\eqref{TcovFS} to relate $\langle \mathcal{C}^0\mathcal{C}^2\rangle$ to $\langle\mathcal{C}^2\mathcal{C}^0\rangle$ and using the Kubo formulas~\eqref{E:kuboTildeChi} we find
\begin{equation}
\label{E:onsagerTildeChi}
\chiB(b_a)-R_0\chiO(b_a) = -\left(\frac{\partial P_0}{\partial\rho_0}\right)_{\!\epsilon_0} \chiE(-b_a) - \left(\frac{\partial P_0}{\partial\epsilon_0}\right)_{\!\rho_0}T\,\chiT(-b_a)\,.
\end{equation}
If we assume that $\chiE$ and $\chiT$ are $\mathbf{T}'$-odd (or alternatively, that $\chiB$ and $\chiO$ are $\mathbf{T}'$-odd), then \eqref{E:onsagerTildeChi} becomes
\begin{equation}
\chiB- R_0 \chiO = \left(\frac{\partial P_0}{\partial\rho_0}\right)_{\!\epsilon_0} \chiE+ \left(\frac{\partial P_0}{\partial\epsilon_0}\right)_{\rho_0}T\,\chiT\,,
\end{equation}
where we have used the fact that the pressure is even under the symmetry $\bf T'$. The same relation follows from the expressions~\eqref{E:newsusceptibilities} that we obtained for the $\tilde{\chi}$'s by demanding the existence of a positive-divergence entropy current (there is also a differential relation between the $\tilde\chi$'s which follows from \eqref{E:newsusceptibilities}).

We have not found any susceptibility or covariance conditions that relate $\etaH$ or $\sigmaH$ to other transport coefficients or thermodynamic functions. However, we do find Kubo formulas for them,
\begin{align}
\label{E:kuboSimple}
\lim_{\omega\rightarrow 0}\frac{G_R^{12,11}(\omega,0)}{i\omega}=\etaH\,,\qquad \lim_{\omega\rightarrow 0}\frac{G_R^{1,2}(\omega,0)}{i\omega}=\sigmaH+\chiE\,,
\end{align}
which can be combined with limits of other retarded functions to give the Kubo formulas~\eqref{E:kubo1} and~\eqref{E:kuboConductivity} presented in Section~\ref{S:summary}.
Finally, the finite-frequency, zero momentum retarded functions $G_R^{12,12}$, $G_R^{11,11}$, and $G_R^{1,1}$ are the same as in the parity-preserving case. The constraints from positivity of the spectral functions are then unchanged, giving~\eqref{E:Restrictions2}.

\section{Thermodynamics}
\label{S:subtractions}
The coefficients $\chiE$, $\chiO$, $\chiB$ and $\chiT$ may be
evaluated from the zero-frequency limits of the retarded functions of
the energy-momentum tensor and the current by \eqref{E:kuboTildeChi}.
The retarded functions at
zero frequency coincide with the Euclidean time-ordered functions at
zero frequency, and the latter can be computed using the Euclidean
functional integral method in the grand canonical ensemble. Therefore,
it seems natural that the coefficients $\chiE$, $\chiO$, $\chiB$ and
$\chiT$ should have a thermodynamic interpretation in terms of the
derivatives of the partition function with respect to the parameters
which leave the system in equilibrium. In this Section we will give such
a thermodynamic interpretation to the functions $\MB(T,\mu)$ and
$\MO(T,\mu)$ which determine the off-equilibrium entropy
current, by treating $B$ and $\Omega$ as parameters which characterize
the equilibrium state. Equations \eqref{E:newsusceptibilities} then
determine the coefficients $\chiE$, $\chiO$, $\chiB$ and $\chiT$ in
terms of the derivatives of the pressure with respect to $B$ and $\Omega$.

Equilibrium states exist with non-zero magnetic field $B$, in which the pressure
will depend on $T$, $\mu$, and $B$.
In addition, on compact manifolds one may have equilibrium configurations with non-zero vorticity $\Omega$. Consider a uniformly rotating fluid on a flat disk with radius $R$ and constant angular velocity $\omega$, satisfying $\omega R \ll 1$. Working to linear order in $\omega R$ the velocity field is given by
\begin{equation}
\label{E:velocity} 	u^{\mu} = \left(1,- \omega x^2,\, \omega x^1 \right) \, ,
\end{equation}
where $x^1$ and $x^2$ are the two spatial coordinates, and the coordinate frame is assumed to be inertial, i.e. $g_{\mu\nu} = \eta_{\mu\nu}= {\rm diag}(-1,1,1)$. The vorticity for this configuration is constant,%
\footnote{\label{F:vort}
	Instead of considering a rotating fluid on a disk, one could consider a
   rotating spacetime in which the fluid velocity is $u^{\mu}=(1,0,0)$
   which is related to \eqref{E:velocity} via a coordinate transformation.
   The value of the vorticity will, of course, be the same for both configurations. 
}
\begin{equation} 	
    -\epsilon^{\mu\nu\rho}u_{\mu}\partial_{\nu} u_{\rho} = 2\omega\,.
\end{equation}
The rotating fluid~\eqref{E:velocity} satisfies the
conservation equations~\eqref{eq:hydro1} with constant energy and charge
densities and exhibits vanishing $\nabla_{\mu}u^{\mu}$, $V_{\mu}$, and $\sigma^{\mu\nu}$. To first order in derivatives the amount of entropy produced is determined by \eqref{E:divs} and \eqref{E:divJscSO2},
\begin{equation*}
\partial_{\mu}J_s^{\mu} = \frac{\zeta (\partial{\cdot} u)^2+\eta \sigma^{\mu\nu}\sigma_{\mu\nu}+\sigma V_{\mu}V^{\mu}}{T}\,.
\end{equation*}
Since $\partial_{\mu}J_s^{\mu} = 0$ for the flow described above we expect it to correspond to an equilibrium state. Other examples of equilibrium configurations with non-zero vorticity can be found, for example, in \cite{bllm,Leigh:2011au}.

If vorticity is a parameter which characterizes an equilibrium state, the thermodynamic pressure will depend on $\mu$, $T$, $B$, and $\Omega$, so that
\begin{align}
dP &= s\,dT + \rho\, d\mu + \frac{\partial P}{\partial B}dB + \frac{\partial P}{\partial \Omega}d\Omega\,, \\
\epsilon + P &= s T + \rho\mu\,.
\end{align}
To first order in $B$ and $\Omega$, the energy-momentum tensor and charge current in such an equilibrium state must take the general form \begin{align}
\label{E:equilibriumTJ}
\begin{split}
    T^{\mu\nu} &= \left(\epsilon - e_B B - e_\Omega \Omega\right)u^{\mu}u^{\nu} + \left(P-\tilde{x}_B B - \tilde{x}_\Omega \Omega\right) \Delta^{\mu\nu}\,, \\
    J^{\mu} & = \left(\rho - r_B B - r_{\Omega} \Omega \right)u^{\mu}\,.
\end{split}
\end{align}
where $e_B$, $e_\Omega$, $\tilde{x}_B$, $\tilde{x}_\Omega$, $r_B$, $r_\Omega$ are functions of $T$ and $\mu$.
Since \eqref{E:equilibriumTJ} describes an equilibrium configuration, the quantities $e_B$, $e_\Omega$, $\tilde{x}_B$, $\tilde{x}_\Omega$, $r_B$, $r_\Omega$ are all measurable. In the remainder of this section we obtain the most general expression for these parameters within the framework described in the previous sections.
Indeed, the expressions \eqref{E:equilibriumTJ} fall into the class of energy-momentum tensors and currents described by \eqref{E:Constitutive2},
where we now view the terms in $\tau^{\mu\nu}$ and $\Upsilon^\mu$ proportional to $B$ and $\Omega$ as equilibrium quantities.
Therefore, we are free to use our results from Sections~\ref{S:entropy} and \ref{S:linearized} specialized to equilibrium configurations with non-zero $B$ and~$\Omega$ in our analysis.

Since we are assuming that $B$ and $\Omega$ label equilibrium states, the Euclidean partition function $Z$ will depend on $\mu$, $T$, $B$, and $\Omega$. 
Recalling footnote \ref{F:vort}, it is convenient to work in a coordinate system where $u^i=0$ in equilibrium. To linear order in fluctuations, the vorticity is  
$\Omega = \partial_1 u^2 - \partial_2 u^1 + \epsilon^{ij} \partial_i h_{0j}$. However, for this equilibrium state and in this coordinate system, $\partial_i u^j$ is higher order in $k$ and the vorticity reduces to $\Omega = \epsilon^{ij}\partial_i h_{0j}$. In this case, since small shifts of the magnetic field $B=\epsilon^{ij} \partial_i A_j$ and the metric perturbations $\epsilon^{ij}\partial_i h_{0j}$ do not affect the radius of the Euclidean time circle or the Wilson line around it, $T$ and $\mu$ should not change to first order in $B$ and $\Omega$. Thus, using the same formalism as the one described in Section \ref{S:GreenGeneral} which led to~\eqref{eq:susceptibility-constraints-1}, we expect that
\begin{subequations}
\label{E:equilibriumG}
\begin{align}
\label{E:susceptMag1}
    \lim_{k \to 0}\frac{1}{i k} G_R^{0,2}(0,k) &= \frac{\partial \rho}{\partial B} - r_B\,, &
    \lim_{k \to 0}\frac{1}{i k} G_R^{00,2}(0,k) &= \frac{\partial \epsilon}{\partial B} - e_B\,, \\
\label{E:susceptVort1}
    \lim_{k \to 0}\frac{1}{i k} G_R^{0,02}(0,k) &= \frac{\partial \rho}{\partial \Omega} - r_{\Omega}\,, &
    \lim_{k \to 0}\frac{1}{i k} G_R^{00,02}(0,k) &= \frac{\partial \epsilon}{\partial \Omega} - e_{\Omega}\,,
\end{align}
\end{subequations}
where $r_B$, $r_\Omega$, $e_B$, and $e_\Omega$ were defined in \eqref{E:equilibriumTJ}.

Now we recall the Onsager relation~\eqref{E:onsagerTildeChi} obtained at the end of Section~\ref{S:linearized}. Combined with the expressions for the $\tilde{\chi}$'s in~\eqref{E:newsusceptibilities}, obtained via the entropy current, we find that the combination $(\partial P_0/\partial\rho_0) \chiE + (\partial P_0/\partial\epsilon_0) T\chiT$ is odd under the symmetry $\mathbf{T}'$, which is time-reversal $\mathbf{T}$ combined with sending the $\mathbf{T}$-violating parameters $b_a \rightarrow - b_a$. In what follows we assume that both $\chiE$ and $\chiT$ are $\mathbf{T}'$-odd.

Applying the Onsager relations~\eqref{TcovFS} to the Kubo formulas~\eqref{E:kuboTE} for $\chiE$ and $\chiT$ we find
\begin{subequations}
\label{E:KuboIntermediate}
\begin{align}
\chiE &= \lim_{k\rightarrow 0}\frac{1}{ik}\left(G_R^{0,2}(0,k) - R_0 G_R^{0,02}(0,k)\right)\,,& T\chiT&=\lim_{k\rightarrow 0}\frac{1}{ik}\left(G_R^{00,2}(0,k)-R_0G_R^{00,02}(0,k)\right)\,.
\end{align}
The Kubo formulas for $\tilde{\chi}_B$ and $\tilde{\chi}_{\Omega}$ in \eqref{E:kuboTB} are also linear combinations of the same retarded functions,
\begin{align}
\begin{split}
\chiB &=\lim_{k\rightarrow 0}\frac{1}{ik}\left(\frac{\partial P_0}{\partial\rho_0}G_R^{0,2}(0,k)+\frac{\partial P_0}{\partial\epsilon_0}G_R^{00,2}(0,k)\right), \\
\chiO&=\lim_{k\rightarrow 0}\frac{1}{ik}\left(\frac{\partial P_0}{\partial\rho_0}G_R^{0,02}(0,k)+\frac{\partial P_0}{\partial\epsilon_0}G_R^{00,02}(0,k)\right).
\end{split}
\end{align}
\end{subequations}
Solving \eqref{E:KuboIntermediate} for $G_R^{0,2}$, $G_R^{00,2}$, $G_R^{0,02}$, and $G_R^{00,02}$ and using~\eqref{E:newsusceptibilities}
we obtain
\begin{subequations}
\label{E:susceptMagVort}
\begin{align}
\label{E:susceptMag2}
\lim_{k\rightarrow 0}\frac{1}{ik}G_R^{0,2}(0,k) &= \frac{\partial\MB}{\partial\mu}\,, \\
\lim_{k\rightarrow 0}\frac{1}{ik}G_R^{00,2}(0,k) &= T\frac{\partial \MB}{\partial T}+\mu\frac{\partial\MB}{\partial\mu}-\MB \,,\\
\label{E:susceptVort2}
\lim_{k\rightarrow 0}\frac{1}{ik}G_R^{0,02}(0,k) &= \frac{\partial \MO}{\partial\mu}-\MB \,, \\
\lim_{k\rightarrow 0}\frac{1}{ik}G_R^{00,02}(0,k) &= T\frac{\partial \MO}{\partial T}+\mu\frac{\partial\MO}{\partial\mu}-2\MO+f_{\Omega}(T) \,.
\end{align}
\end{subequations}
Comparing \eqref{E:susceptMagVort} to \eqref{E:equilibriumG} we find
\begin{subequations}
\label{E:Relations1}
\begin{align}
    T \frac{\partial \MB}{\partial T} + \mu \frac{\partial \MB}{\partial \mu} - \MB &= \frac{\partial \epsilon}{\partial B} - e_B\,, \\
    \frac{\partial \MB}{\partial \mu} &= \frac{\partial \rho}{\partial B} - r_B\,, \\
    T \frac{\partial \MO}{\partial T} + \mu \frac{\partial \MO}{\partial \mu} - \MO &= \frac{\partial \epsilon}{\partial \Omega} +\MO - f_{\Omega}(T) - e_\Omega\,, \\
    \frac{\partial \MO}{\partial \mu} & = \frac{\partial \rho}{\partial \Omega} + \MB - r_\Omega\,.
\end{align}
\end{subequations}
By defining $\bar{r}_{\Omega} = r_{\Omega} - \MB$ and $\bar{e}_{\Omega} = e_{\Omega} - \MO + f_{\Omega}$, equations \eqref{E:Relations1} can be brought to a more symmetric form. We can solve for $\MB$, $\MO$, $e_B$ and ${\bar e}_{\Omega}$ in terms of the functions $r_{B}$ and ${\bar r}_{\Omega}$,
\begin{align}
\begin{split}
\label{E:generalsolutionMMO}
    e_B & = T^2 \frac{\partial}{\partial T} \left(\frac{1}{T}\int^{\mu} r_B d\mu\right) + \mu r_B - T^2 q_B'(T)\,, \\
    {\bar e}_\Omega & = T^2 \frac{\partial}{\partial T} \left(\frac{1}{T}\int^{\mu} \bar{r}_\Omega d\mu \right)+ \mu \bar{r}_\Omega - T^2 q_\Omega'(T)\,, \\
    \MB &= \frac{\partial P}{\partial B} - \int^\mu r_B d\mu + T q_B(T)\,, \\
    \MO &= \frac{\partial P}{\partial \Omega} + \int^\mu {\bar r}_\Omega d\mu + T q_\Omega(T)\,,
\end{split}
\end{align}
where $q_B(T)$ and $q_\Omega(T)$ are undetermined functions.

The coefficients $\tilde{x}_B$ and $\tilde{x}_\Omega$ can be determined from the retarded functions
\begin{align}
\label{E:tiiSuscept}
     \lim_{k\rightarrow 0}\frac{1}{ik}G_R^{ii,2}(0,k) &= \frac{\partial P}{\partial B}-\tilde{x}_B\,, &     \lim_{k\rightarrow 0}\frac{1}{ik}G_R^{ii,02}(0,k) &= \frac{\partial P}{\partial \Omega}-\tilde{x}_\Omega\,,
\end{align}
with no summation over $i$. Using \eqref{E:tii} and \eqref{E:tiiSuscept} we find that
\begin{align}
    \tilde{x}_B = \frac{\partial P}{\partial B}, \qquad   \tilde{x}_{\Omega} = \frac{\partial P}{\partial \Omega}\,.
\end{align}

If $B$ and $\Omega$ are indeed equilibrium parameters, we should find that in an equilibrium state characterized by $T$, $\mu$, $B$ and $\Omega$,
\begin{align}
\label{E:JsArgument}
    u_{\mu} J_s^{\mu} &= -s = -s_0 - \frac{\partial s}{\partial B}B - \frac{\partial s}{\partial \Omega}\Omega\, .
\end{align}
Going back to our expression for the entropy current in \eqref{E:generalEntropyCurrent} and using the expressions for $\Upsilon^\mu$ and $\tau^{\mu\nu}$ from \eqref{E:Constitutive2} and \eqref{E:equilibriumTJ}, we find that \eqref{E:JsArgument} implies
\begin{align}
\label{E:MEntropy}
    T\tilde{\nu}_4 &= \frac{\partial P}{\partial B} + (e_B - \mu r_B )\,, \\
\label{E:n5Entropy}
    T \tilde{\nu}_5 &= \frac{\partial P}{\partial \Omega} + (e_\Omega - \mu r_\Omega )\,.
\end{align}
Remembering the definition $\MB=T\tilde{\nu}_4$ and comparing \eqref{E:MEntropy} with \eqref{E:generalsolutionMMO} we obtain
\begin{equation}
    \frac{\partial}{\partial T}\int^\mu \!r_B\, d\mu = \partial_T(q_B T) \, ,
\end{equation}
from which we conclude that $r_B =r_B(\mu)$ and $q_B(T) = q_0/T$, with constant $q_0$. Defining $h_B(\mu) = q_0 -\int^{\mu} r_B(\mu)d\mu$ we find, for $\Omega=0$, that the equilibrium energy-momentum tensor and current take the form,
\begin{align}
\label{E:equilibriumTJB}
    T^{\mu\nu} &= \left(\epsilon +(\mu h_B'-h_B) B \right)u^{\mu}u^{\nu} + \left(P-\frac{\partial P}{\partial B} B \right) \Delta^{\mu\nu}\,, \\
    J^{\mu} & = \left(\rho + h_B' B \right)u^{\mu}\,,
\end{align}
and
\begin{equation}
    \MB = \frac{\partial P}{\partial B} + h_B(\mu)\,.
\end{equation}
The result~\eqref{E:equilibriumTJB} with $h_B=0$ is the canonical  expression for the stress tensor and current of a fluid at non-zero $B$ and $\Omega=0$, see for example \cite{Cooper:99aa, Hartnoll:2007ih}. Following the literature, we set $h_B=0$, although we have not found a constraint that eliminates it. Note that $h_B=0$
implies that $e_B=r_B=0$.  Then
the only effect of the magnetic field is to shift the equilibrium value of $\frac12\Delta_{\mu\nu}T^{\mu\nu}$ away from the thermodynamic pressure by the term $-(\partial P/\partial B) B$. We will refer to this difference as a ``subtraction''.

Unfortunately, a similar analysis involving the entropy current in equilibrium cannot be carried out for the coefficients $r_{\Omega}$ and $e_{\Omega}$. 
As emphasized below equation \eqref{E:JS-redundance}, the definition of the entropy current is inherently ambiguous. As a result, fixing $\tilde{\nu}_5$ through \eqref{E:n5Entropy} without additional knowledge of $\tilde{\nu}_3$ or $\tilde{\nu}_1$ does not provide additional constraints on the thermodynamic response parameters.
However, based on the similarity of the equations for $\bar{r}_{\Omega}$, $\bar{e}_{\Omega}$ and $\MO$ in (\ref{E:Relations1}) with those for  $r_B$, $e_B$ and $\MB$, we conjecture that
\begin{equation}
    \bar{r}_{\Omega} = \bar{e}_{\Omega}=0\, ,
    \qquad
    \MO = \frac{\partial P}{\partial \Omega}\,.
\end{equation}
This leads us to the equilibrium constitutive relations\,\footnote{Using (\ref{E:expansionS}) and (\ref{E:ToLandau}), we find that the parameters $\tilde{x}_B=\partial P/\partial B = \MB$ and $\tilde{x}_\Omega=\partial P/\partial \Omega = \MO$ are related to $\chiB$ and $\chiO$ in the Landau frame \eqref{E:T1J1L} by
$ \tilde{x}_B = \chiB +
    \left(
    \frac{\partial P}{\partial B} -
    \frac{\partial P_0}{\partial \epsilon_0} \frac{\partial \epsilon}{\partial B} -
    \frac{\partial P_0}{\partial \rho_0} \frac{\partial \rho}{\partial B}
    \right)$, 
    $\tilde{x}_{\Omega} = \chiO +
    \frac{\partial P_0}{\partial \epsilon_0} \left(\MO - f_\Omega \right) +
    \frac{\partial P_0}{\partial \rho_0} \MB +
    \left(
    \frac{\partial P}{\partial \Omega} -
    \frac{\partial P_0}{\partial \epsilon_0} \frac{\partial \epsilon}{\partial \Omega} -
    \frac{\partial P_0}{\partial \rho_0} \frac{\partial \rho}{\partial \Omega}
    \right)$. }
\begin{align}
\begin{split}
\label{E:equilTJwithBO}
    T^{\mu\nu} &= \left(\epsilon - \left(\MO - f_{\Omega} \right)\Omega \right)u^{\mu}u^{\nu} + \left(P-\left(  \MB B +\MO \Omega \right) \right) \Delta^{\mu\nu}\,, \\
    J^{\mu} & = \left(\rho - \MB \Omega \right)u^{\mu}\,,
\end{split}
\end{align}
together with
\begin{equation}
    \MO = \frac{\partial P}{\partial \Omega}\, ,
    \qquad
    \MB = \frac{\partial P}{\partial B}\,.
\end{equation}
These relations determine the `magnetovortical' frame (\ref{E:T1J1}), and we will discuss the above results  in Section~\ref{S:discussion}.

\section{A holographic model}
\label{S:holographic}
\noindent
In this section we obtain explicit expressions for the constitutive relations,
transport coefficients, and thermodynamic response parameters
of parity-violating hydrodynamics in 2+1 dimensions in a relatively simple
model. This is made tractable by the AdS/CFT correspondence.
Our holographic setup is described by the action
\begin{equation}
\label{E:action}
	S = \frac{1}{2\kappa^2}\int \!\!d^4x\,\sqrt{-g}\left[ R+\frac{6}{L^2} - \frac{1}{4}F^2-\frac{1}{2}(\partial\varphi)^2 - V[\varphi] \right]-\frac{1}{64\pi^2}\int d^4x \,\theta[\varphi] \epsilon^{abcd}F_{ab}F_{cd}+S_{\rm bdy},
\end{equation}
where $\theta[\varphi]=\theta_0 + \theta_1 \varphi + \mathcal{O}(\varphi^2)$ is a function specifying the parity-violating axion coupling, the potential satisfies $V[0]=V'[0]=0$, and $S_{\rm bdy}$ is an appropriate boundary contribution to the action. We will denote the coordinates as $x^a=(x^{\mu},r)$, where $r$ foliates the spacetime such that $r\to\infty$ is an asymptotically AdS boundary, and $x^{\mu}$ are coordinates on a constant $r$ hypersurface. We denote the completely antisymmetric Levi-Civita symbol by $\epsilon^{abcd}$ where $\epsilon^{0123}=1$.

The AdS/CFT correspondence~\cite{Maldacena:1997re,Gubser:1998bc,Witten:1998qj} putatively relates the gravitational theory defined by (\ref{E:action})
to a strongly coupled $2+1$-dimensional theory which can be thought of as living on the asymptotically AdS boundary of the spacetime located at $r\to\infty$.
Equation \eqref{E:action} defines a simple two-derivative action which can describe parity violation in the charged sector of the dual field theory. There are a number of string-theoretic embeddings of (\ref{E:action}), including the worldvolume theory of probe branes dual to flavor multiplets~\cite{Bergman:2010gm}, as well as truncations of $11$-dimensional supergravity~\cite{Gauntlett:2009bh}. A similar holographic model, with parity violation introduced through a gravitational axion rather than an electromagnetic axion, has recently 
been studied in~\cite{Saremi:2011ab}.

The action~(\ref{E:action}) leads to the equations of motion:
\begin{align}
\nonumber
R_{ab}-\frac{R}{2}g_{ab}  &= \frac{3}{L^2} g_{ab}+\tau_{ab}\,, \\
\label{eoms}
\partial_{a}\left(\sqrt{-g}F^{ab}\right) &= \frac{\kappa^2}{8\pi^2}\epsilon^{bcde}\partial_{c}\theta F_{de}\,, \\
\nonumber
\partial_{a}\left(\sqrt{-g}\partial^{a}\varphi\right)  &= {\sqrt{-g}}V'[\varphi] + \frac{\theta'[\varphi]\kappa^2}{32\pi^2}\epsilon^{abcd}F_{ab}F_{cd}\,,
\end{align}
where we defined
\begin{equation}
\tau_{ab}=\frac{1}{2}\left[ F_{ac}F_{b}^{\,\,c} - \frac{1}{4}g_{ab}F^2 + \partial_{a}\varphi\partial_{b}\varphi - \frac{1}{2}g_{ab}(\partial\varphi)^2 -g_{ab} V[\varphi]\right]\,,
\end{equation}
which is not to be confused with $\tau^{\mu\nu}$ from \eqref{E:Constitutive2}.
In the remainder of this Section we choose units in which $L=1$.

The AdS/CFT correspondence relates solutions of \eqref{eoms} to expectation values
of operators in the dual $2+1$-dimensional field theory. In particular, the gauge field $A_{\mu}$ is dual to a conserved current $J^{\mu}$ whose expectation value is given by
\begin{equation}
\label{bdyJ}
\langle J^{\mu}(x)\rangle = \frac{1}{\sqrt{-g_{\scriptscriptstyle B}}}\frac{\delta S}{\delta A_{\mu}(x)} = \frac{1}{\sqrt{-g_{\scriptscriptstyle B}}} \lim_{r\rightarrow\infty}\left(-\frac{\sqrt{-g}F^{3\mu}(r)}{2\kappa^2} +\frac{\theta}{16\pi^2}\epsilon^{\mu\nu\rho 3}F_{\nu\rho}(r)\right),
\end{equation}
where $g_{\scriptscriptstyle B}$ is the background metric of the dual field theory. Similarly, the scalar field $\varphi$ is dual to a neutral scalar operator $\mathcal{O}_{\varphi}$. The conformal dimension $\Delta$ of $\mathcal{O}_{\varphi}$, satisfying $3/2 < \Delta < 3$,\footnote{
Unitarity implies that $\Delta \geq 1/2$, while ${\mathcal O}_\varphi$ is a relevant operator if $\Delta \leq 3$.
We focus here on $\Delta \in (3/2,3)$ for computational convenience. We expect that our results may be analytically continued to $1/2 \leq \Delta \leq 3$.} is related to the mass of $\varphi$ through $V''[0] = \Delta (\Delta-3)$. We will break conformal invariance by turning on a source term $J_{\varphi} = \Lambda^{3-\Delta}$ dual to $\mathcal{O}_{\varphi}$,
with $\Lambda$ characterizing the strength of the source.
In other words, we will deform the field theory action by a $\frac{1}{\Delta-\frac{3}{2}} \int\! d^3x \sqrt{-g_{\scriptscriptstyle B}} J_{\varphi} \mathcal{O}_{\varphi}$ term, where
the prefactor $(\Delta-\frac32)^{-1}$ was introduced in \cite{Klebanov:1999tb} in order to have a smooth $\Delta \to 3/2$ limit.
Near the boundary, we find according to \cite{Klebanov:1999tb} that
\begin{equation}
\label{bdyPhi}
\varphi(r{\to}\infty) = \frac{\Lambda^{3-\Delta}}{\Delta-\frac{3}{2}}\, r ^{\Delta- 3}\left(1+\mathcal{O}(r^{-1})\right) +\frac{\kappa^2\langle \mathcal{O}_{\varphi}\rangle}{\Delta-\frac32}\, r^{-\Delta}\left(1+\mathcal{O}(r^{-1})\right)\,.
\end{equation}
We will restrict our attention to the high-temperature regime $T\gg \Lambda$, $T\gg\mu$ which evades potential low-temperature instabilities, and work perturbatively in small $\mu$, $J_\varphi$, $B$, and $\Omega$.
The expectation value of the energy-momentum tensor of the boundary theory can be computed along similar lines to \cite{Balasubramanian:1999re,Bianchi:2001de}. However, in this work we  will only need the trace $T_\mu^\mu$ of the boundary theory which can be computed from $\mathcal{O}_{\varphi}$ using
\begin{equation}
\label{wardTrace}
\langle T_{\mu}^{\mu}\rangle = \frac{3-\Delta}{\Delta-\frac{3}{2}}\Lambda^{3-\Delta}\langle\mathcal{O}_{\varphi}\rangle.
\end{equation}
The Ward identity \eqref{wardTrace} can be derived holographically as in \cite{Bianchi:2001de}, or by a field theory computation as in, for example, \cite{Petkou:1999fv}. For particular values of $\Delta$ there may be an anomalous contribution to the trace of the stress tensor associated with a matter conformal anomaly. We study values of $\Delta$ for which this anomaly is absent.

\subsection{Equilibrium configurations}
In this subsection we construct equilibrium solutions to \eqref{eoms}
to first order in the magnetic field and vorticity.
Various solutions to (6.2) with $\varphi=0$ and non-vanishing magnetic field and vorticity do appear in the literature~\cite{Romans:1991nq,Carter:1968ks,Plebanski:1976gy,Caldarelli:1999xj}. However, a more useful starting point for our analysis is the solution described, for example, in \cite{Yarom:2009mw} (see also \cite{Hohler:2009tv,Cherman:2009tw}),
\begin{align}
\nonumber
g_{ab} dx^{a}dx^{b} &= 2 dt \,dr + r^2(-f(r)dt^2 + d\mathbf{x}^2)\, , \hspace{.5in} f(r) = 1 - \frac{{r_H}^3}{r^3} + \mathcal{O}(\mu^2,\,J_{\varphi}^2) \,, \\
\label{pertBG}
A_a dx^{a} & = \left( \mu - \frac{q}{r}\right)dt+ \mathcal{O}(J_{\varphi}^2)\,, \\
\nonumber
\varphi & =  \left(\frac{\Lambda}{{r_H}}\right)^{3-\Delta}\frac{2 \Gamma\left(\frac{\Delta}{3}\right)^2}{\Gamma\left( \frac{2\Delta}{3} \right)} P_{\frac{\Delta}{3}-1}\!\left( -1 + \frac{2r^3}{{r_H}^3}\right) + \mathcal{O}(\mu^2,\,J_{\varphi}^3) \,,
\end{align}
where $P_\lambda(z)$ is a Legendre function of the first kind.
The temperature of the boundary theory is given by
\begin{equation}
\label{E:Trh}
T = \frac{3{r_H}}{4\pi} + \mathcal{O}(\mu^2,\,J_{\varphi}^2)\,,
\end{equation}
the chemical potential is $\mu = q/r_H$, and the charge density is $\rho = q/2\kappa^2$.

We would like to turn on a small magnetic field and vorticity such that the boundary theory stays in equilibrium. Consider the ansatz,
\begin{align}
\begin{split}
\label{pertBO}
	\delta g_{ab} dx^{a}dx^{b}  &= 2 \Omega r^2 f(r)x^1 dt\, dx^2 - \Omega x^1 dr dx^2 + \mathcal{O}(J_{\varphi}^2)\,,  \\
	\quad
	\delta A_a dx^{a} &= \left(B + \frac{\mu {r_H} \Omega}{r}\right)x^1 dx^2 + \mathcal{O}(J_{\varphi}^2)\,,
\end{split}
\end{align}
which describes a configuration in the boundary theory with non-zero vorticity $\Omega$ and non-zero magnetic field $B$.
When $\theta[\varphi]=0$, \eqref{pertBO} solves the equations of motion \eqref{eoms} to linear order in $B$ and $\Omega$. We may now solve \eqref{eoms} for $\theta[\varphi] \neq 0$ by further perturbing  the solution in \eqref{pertBO} to linear order in $B$ and $\Omega$. We find
\begin{align}
\begin{split}
\label{pertSol}
	\delta A_{0} &= -\frac{\kappa^2\theta_1}{4\pi^2}\left[ \frac{{r_H}}{r}\alpha_0+\int_{\infty}^r\frac{dr'}{r'^2}\int_{r_H}^{r'}dr''\,\varphi'(r'')\left(B+\frac{\mu {r_H} \Omega}{r''}\right)\right]
	+\mathcal{O}(\mu^2, \, J_{\varphi}^2)
	\,, \\
	\delta\varphi &= -\frac{\kappa^2\theta_1 \mu {r_H}}{4\pi^2}\varphi(r)\int_{\infty}^r\frac{dr'}{f(r')r'^4\varphi(r')^2}\int_{r_H}^{r'}\frac{dr''\,\varphi(r'')}{r''^2}\left(B+\frac{\mu {r_H} \Omega}{r''}\right)
	+\mathcal{O}(\mu^2 B,\,\mu^3\Omega,\,J_{\varphi})\,, \hspace{-0.4in}
\end{split}
\end{align}
where $\theta_1$ is defined below \eqref{E:action},
and $\alpha_0$ is a  constant, chosen so that $\delta A_{0}$ vanishes at the horizon,
\begin{equation}
\alpha_0=\int_{r_H}^{\infty}\frac{dr}{r^2}\int_{r_H}^{r}dr'\varphi'(r')\left(B+\frac{q\Omega}{r'}\right).
\end{equation}
In obtaining \eqref{pertSol}, we have applied equilibrium boundary conditions to the equations of motion. Namely, regularity at the horizon and normalizability at the AdS boundary. This last condition corresponds to holding the sources of the boundary theory, $\mu = \lim_{r\rightarrow\infty}A_0$ and $J_{\varphi} = \Lambda^{3-\Delta}$, fixed. The perturbations $\delta A_0$ and $\delta\varphi$ also source metric
corrections,
whose effect on susceptibilities is suppressed at order $\mu^2 J_{\varphi}$.
The magnetic field and vorticity appear in the combination $B+q\Omega/r$,
which is a consequence of $\Omega$ appearing in \eqref{pertBO}.

Using~\eqref{bdyJ} and \eqref{bdyPhi} and integrating by parts we find that the boundary theory current and scalar expectation values are shifted by
\begin{subequations}
\begin{align}
\label{deltaQ}
	\delta \langle J^0\rangle =& \frac{\theta_1 {r_H}}{8\pi^2}\left[ B \int_{r_H}^{\infty}\frac{dr\, \varphi}{r^2}+\mu\Omega \left( 2{r_H}\int_{r_H}^{\infty}\frac{dr\,\varphi}{r^3}-\int_{r_H}^{\infty}\frac{dr \,\varphi}{r^2} \right)\right] \nonumber\\
	& + \frac{\theta_0 B}{8 \pi^2} + \mathcal{O}(\mu^2,\mu^2 B,\mu^2\Omega, J_{\varphi}^2)\,, \\
\label{deltaVeV}
	\delta \langle \mathcal{O}_{\varphi}\rangle =&  \Lambda^{\Delta-3}\left(\Delta - \frac{3}{2}\right) \frac{\theta_1\mu {r_H}}{8\pi^2}\int_{r_H}^{\infty}\frac{dr\,\varphi}{r^2}\left( B+\frac{\mu {r_H}\Omega}{r}\right) +\mathcal{O}(\mu^2 B,\,\mu^3\Omega,\,J_{\varphi})\,.
\end{align}
\end{subequations}
In what follows it will be convenient to define
\begin{equation}
	\Phi_n = \frac{\theta_1 {r_H}}{8 \pi^2} \int_{r_H}^{\infty} \!dr\,\frac{\varphi(r)}{r^n}\,.
\end{equation}
where $\varphi(r)$ is given by \eqref{pertBG}. Using \eqref{wardTrace} we infer that, at $B=\Omega =0$,
\begin{subequations}
\begin{align}
\label{drho}
	\frac{\partial \langle \delta J^0\rangle}{\partial B} &=  \Phi_2 + \frac{\theta_0}{8 \pi^2} +\mathcal{O}(\mu^2,\,J_{\varphi}^2)\,, &
	\frac{\partial \langle \delta J^0\rangle}{\partial \Omega} &= 2 \mu {r_H} \Phi_3 - \mu \Phi_2 + \mathcal{O}(\mu^2,\,J_{\varphi}^2)\,, \\
\label{dTr}
	\frac{\partial\langle \delta T_{\mu}^{\mu}\rangle}{\partial B}&= (3-\Delta)\mu \Phi_2 + \mathcal{O}(\mu^2,\,J_{\varphi}^2)\,,&
	\frac{\partial\langle \delta T_{\mu}^{\mu}\rangle}{\partial \Omega}&= (3-\Delta) \mu {r_H} \Phi_3 + \mathcal{O}(\mu^3,\,J_{\varphi}^2)\,.
\end{align}
\end{subequations}
We are now in a position to compute all the susceptibilities.
In particular, using the expressions $J^0 = \rho - \MB \Omega$, $T^{00} = \epsilon   +(f_{\Omega}(T)-\MO) \Omega$, $T^{11} = T^{22} = P - \MB B - \MO \Omega$ and $dP = s dT + \mu d\rho + \MB dB + \MO d\Omega$ from Section \ref{S:subtractions}, we find
\begin{subequations}
\label{E:susceptibilities}
\begin{align}
\frac{\partial\rho}{\partial B}&=\Phi_2+\frac{\theta_0}{8\pi^2}+\mathcal{O}(\mu^2,J_{\varphi}^2)\,, & \frac{\partial\epsilon}{\partial B} = (\Delta-3)\mu\Phi_2+\mathcal{O}(\mu^2,J_{\varphi}^2)\,.
\end{align}
Integrating the Maxwell relation $\partial\rho/\partial B = \partial \MB/\partial\mu$ leads to
\begin{equation}
\frac{\partial P}{\partial B}=\MB = \mu\Phi_2+\frac{\mu\theta_0}{8\pi^2}+c_B(T)+\mathcal{O}(\mu^2,J_{\varphi}^2)\,,
\end{equation}
where $c_B$ is the magnetization density at $\mu=0$. However, $c_B$ vanishes due to the symmetry of the model under charge conjugation $\mathbf{C}$. Having obtained $\MB$, we find the vortical variation of $\rho$, via $J^0=\rho-\MB\Omega$, to be
\begin{equation}
\frac{\partial\rho}{\partial\Omega}=2\mu r_H\Phi_3+\frac{\mu\theta_0}{8\pi^2}+\mathcal{O}(\mu^2,J_{\varphi}^2)\,.
\end{equation}
We then integrate the Maxwell relation $\partial\rho/\partial\Omega =\partial\MO/\partial\mu$ to find
\begin{equation}
\frac{\partial P}{\partial\Omega}=\MO = \mu^2r_H \Phi_3+\frac{\mu^2\theta_0}{16\pi^2}+c_\Omega(T) +\mathcal{O}(\mu^3,J_{\varphi}^2)\,,
\end{equation}
where $c_\Omega$ is $\partial P/\partial\Omega$ at $\mu=0$. We now point out that at $B=0$ and $\mu=0$ the perturbation~\eqref{pertBO} is an exact solution to~\eqref{eoms} to linear order in $\Omega$. It follows that the on-shell bulk action density, equal to the pressure of the boundary theory, is unchanged at this order so that $\partial P/\partial\Omega$ vanishes at $\mu=0$. This eliminates $c_\Omega$. Using the expressions for $T^{00}$ and $T^{ii}$, we find the final susceptibility to be
\begin{equation}
\frac{\partial\epsilon}{\partial\Omega}=(\Delta-2)\mu^2r_H\Phi_3-f_{\Omega}(T)+\mathcal{O}(\mu^3,J_{\varphi}^2)\,.
\end{equation}
\end{subequations}

We can determine $f_{\Omega}(T)$ by using
\begin{equation}
\label{TdTphi}
T\partial_T\varphi = {r_H}\partial_{r_H} \varphi = (\Delta - 3)\varphi - r\varphi'\,,
\end{equation}
which follows from \eqref{pertBG}, to show that the magnetic and vortical variations of $\epsilon$ are
\begin{subequations}
\label{E:dTrBO}
\begin{align}
\label{dtrdB}
	 \frac{\partial\epsilon}{\partial B} &= T \frac{\partial\MB}{\partial T} + \mu\frac{\partial\MB}{\partial\mu} - \MB\,, \\
\label{dtrdO}
	 \frac{\partial\epsilon}{\partial\Omega} &= T \frac{\partial \MO}{\partial T} + \mu \frac{\partial \MO}{\partial \mu} - \MO-f_{\Omega}(T)\,,
\end{align}
\end{subequations}
at the order to which we are working. The thermodynamic relations~\eqref{dtrdB} and~\eqref{dtrdO} follow from~\eqref{E:firstLaw}, $dP = s dT + \rho d\mu + \MB dB + \MO d\Omega$, and~(\ref{E:eos2}), $\epsilon + P = s T + \mu \rho$, only when $f_{\Omega}(T)=0$, implying that $f_\Omega$ vanishes. We emphasize that in obtaining~\eqref{E:susceptibilities} we used~\eqref{E:equilTJwithBO} without deriving it explicitly in the holographic model. However, we note that (\ref{E:dTrBO}) represents a nontrivial consistency check of (\ref{E:equilTJwithBO}).

\subsection{Fluid-gravity correspondence}
\noindent
The fluid-gravity correspondence \cite{Bhattacharyya:2008jc} provides a method to explicitly compute the hydrodynamic constitutive relations of the boundary theory.
In this subsection we will use it to check the consistency of the parity-violating constitutive relations described in \eqref{E:T1J1}, \eqref{E:susceptibilitiesMV} and \eqref{E:susceptibilities}.
We begin with the background (\ref{pertBG}) boosted to a velocity  $u^{\mu}$ normalized so that $u^{\mu}\eta_{\mu\nu} u^{\nu}=-1$, with $\eta$ the Minkowski metric. In what follows we define $u_{\mu} = \eta_{\mu\nu}u^{\nu}$ and
\begin{align}
\begin{split}
\label{pertBGu}
g_{ab}dx^a dx^b&= -2 u_{\mu}dx^{\mu}dr + r^2(-f(r)u_{\mu}u_{\nu}+\Delta_{\mu\nu})dx^{\mu}dx^{\nu} + \mathcal{O}(\mu^2,\,J_{\varphi}) \,, \\
A_a dx^a &=  \frac{\mu {r_H}}{r}u_{\mu}dx^{\mu} + \mathcal{O}(\mu^2,\,J_{\varphi}^3)\,,  \\
\varphi & =  \left(\frac{\Lambda}{{r_H}}\right)^{3-\Delta}\frac{2 \Gamma\left(\frac{\Delta}{3}\right)^2}{\Gamma\left( \frac{2\Delta}{3} \right)} P_{\frac{\Delta}{3}-1}\!\left( -1 + \frac{2r^3}{{r_H}^3}\right) + \mathcal{O}(\mu^2,\,J_{\varphi}^2)\, ,
\end{split}
\end{align}
with $f(r)$ as in \eqref{pertBG}.
We would like to retain the distinction between a chemical potential and an external background field in an out-of-equilibrium configuration. For this reason, the gauge field in~\eqref{pertBGu}
has been shifted relative to the one in~\eqref{pertBG}. When $u_\mu$, $T$, and $\mu$ are constant, the profiles \eqref{pertBGu} solve the equations of motion \eqref{eoms}.

Consider an extension of the solution \eqref{pertBGu} where $\mu$, $T$ and $u^{\mu}$ vary slowly as a function of the $x^{\mu}$ coordinates, and, in addition, an external electromagnetic field whose magnitude is of the order of the gradients of $\mu$, $T$ and $u^{\mu}$, is turned on. This extension of (\ref{pertBGu}) no longer solves the equations of motion (\ref{eoms}), so the bulk fields need to be corrected. Since we have chosen $u^{\mu}$, $T$ and $\mu$ to depend slowly on the spacetime coordinates, we can correct the bulk fields order by order in derivatives of $u^{\mu}$, $T$ and $\mu$.
We write
\begin{equation}
g = g^{(0)} + g^{(1)}+ \cdots, \hspace{.3in} A = A^{(0)}+A^{(1) }+ \cdots, \hspace{.3in} \varphi=\varphi^{(0)} + \varphi^{(1)}+\cdots \,,
\end{equation}
where $g^{(0)}$, $A^{(0)}$, and $\varphi^{(0)}$ refer to \eqref{pertBGu} with slowly varying $T$, $\mu$, and $u_\mu$. See~\cite{Bhattacharyya:2008jc,Banerjee:2008th,Erdmenger:2008rm} for further details of the derivative expansion.

We parametrize the corrections to the metric and gauge field by
\begin{align}\nonumber
g^{(1)}_{ab}dx^{a}dx^{b} &= -2g_1(r) u_\mu dx^{\mu}dr+r^2(g_0(r)u_{\mu}u_{\nu}+ \gamma_{\mu}u_{\nu } + \gamma_{\nu}u_{\mu} +\pi_{\mu\nu}+g_1(r)\Delta_{\mu\nu})dx^{\mu}dx^{\nu}\,, \\
A^{(1)}_a dx^{a} &= -a_0(r)u_{\mu}dx^{\mu}+\alpha_{\mu}(r) dx^{\mu}+A_{\mu}^{\rm bg}dx^{\mu}\,,
\end{align}
where $\alpha_{\mu}(r)$ and $\gamma_{\mu}(r)$ are transverse to $u^{\mu}$, $\pi_{\mu\nu}(r)$ is symmetric, traceless, and transverse to $u^{\mu}$, and $A_{\mu}^{\rm bg}(x^\lambda)$ is an arbitrary electromagnetic background gauge field, to first order in derivatives.
To first order in the chemical potential, we can decouple the equations for $\gamma_\mu(r)$ and $\alpha_\mu(r)$.
We find
\begin{multline}
\label{E:alpha}
(f r^2 \alpha_{\mu}')' = \frac{\Delta^{\mu\nu}\left(\partial_{\mu}q+ 2qu^{\alpha}\partial_{\alpha} {u}_{\mu}\right)}{r^2} \\
	 - \frac{\kappa^2}{4\pi^2}\theta_1\epsilon^{\mu\nu\rho}u_{\nu}\left[\left( E_{\text{bg},\rho} - \frac{\partial_{\rho}q+qu^{\alpha}\partial_{\alpha} {u}_{\rho}}{r}\right)\varphi^{(0)'}-\frac{q}{r^2}\partial_{\rho}\varphi^{(0)}\right] + \mathcal{O}(\mu^2,J_{\varphi}^2)\,,
\end{multline}
with $q={r_H} \mu$ and $E_{\rm bg}^{\mu} = F^{\mu\nu}_{\rm bg} u_{\nu}$. Equation \eqref{E:alpha}
 may immediately be integrated to obtain $\alpha_{\mu}$ and, via (\ref{bdyJ}), the transverse component of the boundary current. We now translate this result into the constitutive relations given in (\ref{eq:nu1}). To this end, we need to relate $u^{\mu}\partial_{\mu} {u}_{\rho}$ to $\Delta^{\mu\nu}\partial_{\nu}T$, $\Delta^{\mu\nu}\partial_{\nu}\mu$ and $E_{\rm bg}^{\mu}$ using the equations of motion of ideal hydrodynamics. Energy-momentum conservation of the boundary theory is encoded in the Einstein equations, as described in
 \cite{Hur:2008tq}, yielding
\begin{equation}
	u^{\nu}\partial_{\nu} {u}^{\mu} = \frac{\mu E^{\mu}_{\rm bg}}{{r_H}^2} - \frac{\Delta^{\mu\nu}\partial_{\nu} {r_H}}{{r_H}}\,,
\end{equation}
which is just the transverse vector part of the ideal hydrodynamic equations. After using $q=r_H \mu$, \eqref{E:Trh}, \eqref{E:susceptibilities} and \eqref{TdTphi}, we find that the constitutive relations of the holographic fluid take the form~(\ref{E:T1J1}) with
\begin{subequations}\label{E:holoCoeffs}
\begin{align}
\sigma &= \frac{1}{2\kappa^2}+\mathcal{O}(\mu^2,J_{\varphi}^2)\,, &\sigmaH &= \frac{\theta_1}{8\pi^2}\varphi({r_H})  - \frac{\partial\rho}{\partial B}+\mathcal{O}(\mu^2,J_{\varphi}^2)\,, \\
\chiE  &= \frac{\partial\rho}{\partial B}+\mathcal{O}(\mu^2,J_{\varphi}^2)\,, &T\chiT  &= \frac{\partial\epsilon}{\partial B}+\mathcal{O}(\mu^3,J_{\varphi}^2)\,,
\end{align}
\end{subequations}
where the susceptibilities are given in \eqref{E:susceptibilities},
and the Hall viscosity vanishes, $\tilde{\eta} = 0$.\footnote{The linearized equations of motion for the tensor modes of the metric $h_{12}$ and $(h_{11}{-}h_{22})$, which follow from \eqref{E:action}, decouple from the equations for $A_a$ and $\varphi$ at $k=0$. The Kubo formula \eqref{E:kuboSimple} then implies that $\tilde\eta=0$ is an exact relation in this model.}
The constitutive relations with \eqref{E:holoCoeffs} match our earlier predictions (\ref{E:newsusceptibilities}) once we recall that $\MB=\mathcal{O}(\mu)$, $\MO=\mathcal{O}(\mu^2)$, $\partial\rho/\partial\Omega =\mathcal{O}(\mu)$, and $\partial\epsilon/\partial\Omega = \mathcal{O}(\mu^2)$, $R_0=\mathcal{O}(\mu)$, so that their
contributions to $\chiE$ and $\chiT$ are $\mathcal{O}(\mu^2)$ and $\mathcal{O}(\mu^3)$, respectively. We also note that at this order the off-diagonal conductivity is given by
\begin{equation}
	\sigmaH+\chiE = \frac{\theta(\varphi({r_H})) }{8\pi^2}+\mathcal{O}(\mu^2)\, ,
\end{equation}
which is reminiscent of the membrane paradigm~\cite{Thorne:1986iy}.

We may compute the trace of the stress tensor by solving for the correction to the scalar field and  using (\ref{wardTrace}). At high temperatures the scalar equation of motion becomes
\begin{equation}
(f r^4\varphi^{(1)\,\prime})' - r^2\Delta(\Delta-3)\varphi^{(1)} = -\frac{\kappa^2\theta_1 \mu {r_H}}{4\pi^2r^2} \left(B_{\rm bg}+\frac{\mu {r_H} \Omega}{r} \right)+\left( \substack{\hbox{\tiny parity-even} \\ \hbox{\tiny terms} }\right) ,
\end{equation}
where we have used the definitions $B_{\rm bg} = -\epsilon^{\mu\nu\rho}u_{\mu}F^{\rm bg}_{\nu\rho}/2$ and $\Omega = -\epsilon^{\mu\nu\rho}u_{\nu}\partial_{\nu}u_{\rho}$.
The parity-violating part of the solution for $\varphi^{(1)}$ is the same as that for $\delta\varphi$,~\eqref{pertSol}, as described in the previous subsection. Therefore, the  shift in the trace of the energy-momentum tensor coincides with that described in \eqref{dTr},
\begin{equation}
\delta \langle T_{\mu}^{\mu}\rangle =-B\frac{\partial\epsilon}{\partial B} -\Omega \left(\frac{\partial\epsilon}{\partial \Omega} - \MO\right) + \cdots,
\end{equation}
where the omitted terms are independent of $B$ and $\Omega$, and we have used the equilibrium susceptibilities \eqref{E:susceptibilities}. It is satisfying that the magnetic and vortical subtractions computed in equilibrium match the dynamical result as expected on general grounds.

\section{Discussion}
\label{S:discussion}

In this work, we have presented  a general framework for first-order relativistic hydrodynamics in $2+1$ dimensions, applicable to systems with parity violation. We have attempted to describe our computations and results in as simple a form as possible, avoiding  unnecessary interpretation, with the results summarized in Section \ref{S:summary}. In this section we include a discussion of a few properties of this hydrodynamic framework.

\begin{itemize}

\item{\bf Thermodynamic and hydrodynamic response.}
    The Kubo formulas computed in Section~\ref{S:linearized}, and summarized in Section~\ref{S:summary}, naturally split into zero momentum (or hydrodynamic) correlators and zero frequency (or thermodynamic) correlators. The parameters $\eta$, $\zeta$, $\sigma$, $\etaH$, and $\sigmaH$ obtained from zero-momentum correlators are transport coefficients, while $\chiE$, $\chiT$, $\chiB$ and $\chiO$ obtained from zero-frequency correlators are thermodynamic response parameters. The difference between hydrodynamic and thermodynamic response parameters has been discussed previously in the literature. See, for example, \cite{Baier:2007ix,Bhattacharyya:2008jc, Moore:2010bu} for a discussion of thermodynamic response parameters in second order hydrodynamics in $3+1$ dimensions, or \cite{Amado:2011zx} for examples of thermodynamic response parameters in first order parity-violating hydrodynamics in $3+1$ dimensions. Curiously, the Kubo formulas for the anomalous $3+1$-dimensional transport coefficients $\xi_B$ and $\xi$, defined through the Landau frame charge current
    \begin{equation}
        J^{\mu}_{\rm 3+1} = \rho u^\mu + \sigma V^{\mu} + \xi_B B^{\mu}+\xi  w^{\mu},
    \end{equation}
    with $B^{\mu}=\frac12\epsilon^{\mu\nu\rho\sigma}u_{\nu}F_{\rho\sigma}$ the magnetic field and $w^{\mu}=\epsilon^{\mu\nu\rho\sigma}u_{\nu}\nabla_{\rho}u_{\sigma}$ the vorticity, are tantalizingly similar to the Kubo formula \eqref{E:kubo2}  for $\chiE$ and $\chiT$.

\item{\bf Thermodynamic subtractions.}
In Section \ref{S:subtractions} we obtained the equilibrium expressions
for the energy-momentum tensor and the current~\eqref{E:equilTJwithBO},
\begin{align}
\label{E:equilTJwithBO-2}
\begin{split}
    T^{\mu\nu} &= \left(\epsilon - \left(\MO - f_{\Omega} \right) \Omega\right)u^{\mu}u^{\nu} + \left(P-\left(  \MB B +\MO \Omega \right) \right) \Delta^{\mu\nu}\,, \\
    J^{\mu} & = \left(\rho - \MB \Omega \right)u^{\mu}\,.
\end{split}
\end{align}
The above structures can be given a physical interpretation.
First, the magnetic subtraction to the thermodynamic pressure can be understood as being due to the force exerted by $B$ on the boundary currents of the fluid elements~\cite{Cooper:99aa}.
Second, a rotating fluid with non-zero magnetization must also have non-zero
polarization (see for example~\cite{Ridgely-AJP}), thereby inducing
effective ``bound charges.'' In \cite{Hartnoll:2007ih} the latter effect
manifests itself as an extra correction to the current which takes the
form $J^{\mu} = \rho u^{\mu} + \partial_{\nu}M^{\mu\nu}$. Rewriting
$M^{\mu\nu} = - \MB \epsilon^{\mu\nu\rho} u_{\nu}$, taking $\mu
$ and $T$ to be constant and using the equations of motion
\eqref{eq:hydro1} to show that the transverse part of $\epsilon^{\mu\nu
\rho}\partial_{\nu}u_{\rho}$ vanishes, we obtain
$J^\mu = (\rho - \MB \Omega) u^\mu$ as in (\ref{E:equilTJwithBO-2}).

\item{\bf Restrictions on the constitutive relations.}
Had we known the equilibrium stress tensor and current~\eqref{E:equilTJwithBO-2} at non-zero $B$ and $\Omega$, we could have obtained the susceptibility conditions~\eqref{E:susceptMagVort} without requiring any input from the entropy current. Applying \eqref{E:susceptMagVort} and the Onsager relations~\eqref{TcovFS} to the Kubo formulas for the $\tilde{\chi}$'s~\eqref{E:kuboTildeChi} then determines $\MB$, $\MO$, $\chiE$, $\chiT$, $\tilde{x}_B$ and $\tilde{x}_{\Omega}$ to take on the same values we found by demanding positivity of entropy production. It is interesting that the constraints on hydrodynamics from correlation functions then precisely match those obtained from the entropy current. In each case the $\tilde{\chi}$'s are functions of  $\partial P/\partial B$ and $\partial P/\partial \Omega$ and the integration constant $f_{\Omega}(T)$. Note that $\partial P/\partial B$ can be interpreted as the magnetization density while, as discussed below, $\partial P/\partial \Omega$ can be interpreted as half the angular momentum density of the equilibrium state.

\item{\bf Equilibrium states with vorticity.}
The possibility of equilibrium configurations with non-zero vorticity $\Omega$ was discussed in Section \ref{S:subtractions}.
The example we gave involved a limit in which the vorticity was taken to be small, of the same scale as gradients of thermodynamic variables. In this limit, there is a solution to the hydrodynamic equations in which the fluid is rotating with constant angular frequency $\omega$ and has vorticity $\Omega=2\omega$. We point out that a fluid rotating in a flat background is equivalent, via a diffeomorphism, to a non rotating fluid moving at constant velocity, in a rotating background. If we denote $B_G = \epsilon^{ij}\partial_i h_{0j}$, also known as the gravitomagnetic field, we find using the aforementioned diffeomorphism that $B_G = 2\omega$.
    This relation has been used recently in a study of response in topological superconductors and superfluids~\cite{Nomura:2011hn}. Equilibrium states in the presence of gravitomagnetic fields were considered previously in~\cite{2010arXiv1010.0936R}. We observe that diffeomorphism invariance implies that equilibrium states characterized by either a constant angular frequency or a constant gravitomagnetic field are equivalent to states with non-zero vorticity (see also \cite{Amado:2011zx}).
    While the vorticity $\Omega$ is diffeomorphism invariant, $B_G$ and $\omega$ are not.

    There also exist equilibrium configurations where $\Omega$ does not need to be small. One example is given by a fluid on a sphere \cite{bllm} where $\Omega = 2 \omega \cos\theta/(1-\omega^2 \sin^2\theta)$, and $\theta$ is the polar angle on the two-sphere. Another candidate example is an extension of our solution from Section \ref{S:subtractions}, where $u^{\mu} = \gamma (1,\,0,\,\omega)$, $\gamma = 1/\sqrt{1-(\omega r)^2}$, and we are working in a polar $(t,r,\theta)$ coordinate system. In this configuration we find that $\Omega=2\omega \gamma^2$ while the shear and divergence of the velocity field vanish, $\sigma_{\mu\nu}=0$ and $\nabla_{\mu}u^{\mu}=0$. See  \cite{Leigh:2011au} for additional examples.

\item{\bf Vorticity and angular momentum.}
In an equilibrium configuration where either the spatial fluid velocity or angular velocity are constant but non-zero, the partition function \eqref{E:Zdef} takes the form
   \begin{equation}
   \label{E:Zdef2}
      Z[T,\mu,u^{\mu}] = \hbox{Tr}\left[\exp \frac{1}{T}
       \int\! d^2x\left(u_{\mu}T^{0\mu} + {\mu J^0}\right)\right].
   \end{equation}
Inserting the flat space rotating solution \eqref{E:velocity} from Section~\ref{S:subtractions} which has non-zero vorticity $\Omega$, together with the definition $L=\int\! d^2x\,\epsilon_{ij}\, x^i T^{0j}$ for the total angular momentum, leads to
  \begin{equation}
  \label{E:Zdef3}
     Z = \hbox{Tr}\left[\exp\left(-\frac{H}{T} + \frac{L \Omega}{2 T} + \frac{\mu Q}{T}\right)\right]\,.
  \end{equation}
We observe that \eqref{E:Zdef3} describes an ensemble with non-zero total angular momentum.

\item{\bf Ferromagnetism and Ferrovorticism.} When $\Omega$ characterizes equilibrium states, an interesting feature of {\bf P}-violating systems is that the response of the pressure to vorticity, $\MO$, and to a magnetic field, $\MB$, can be non-zero when $\Omega{=}0$ and $B{=}0$. While a non-zero value of magnetization density $\MB=\partial P/\partial B$ at $B{=}0$ is associated with ferromagnetism, one may term a similar phenomenon of non-zero $\MO=\partial P/\partial \Omega$ at $\Omega{=}0$ ``ferrovorticism''. For the rigid rotation states \eqref{E:velocity} discussed in Section~\ref{S:subtractions} with constant vorticity $\Omega=2\omega$, the form of the partition function (\ref{E:Zdef3})
shows that we have
    \begin{equation}
    \MO = \left.\frac{\partial P}{\partial \Omega}\right|_{\Omega=0} = \frac{1}{2} \left.\frac{\partial P}{\partial \omega}\right|_{\omega=0} = \frac{1}{2} \langle {\ell} \rangle\,,
    \end{equation}
where $\langle \ell \rangle=\langle L\rangle/{\cal V}$ denotes the density of the total angular momentum at zero rotation frequency. In parity-invariant fluids, the total angular momentum in the parity-invariant thermal equilibrium state with $\omega{=}0$ would have to vanish. However, when parity is not a symmetry of the microscopic theory, there is no reason why $\langle L \rangle$ should vanish at $\omega{=}0$ (just as there is no reason that the total charge $\langle Q\rangle$ should vanish at $\mu{=}0$ when charge conjugation is not a symmetry). The holographic model studied in Section~\ref{S:holographic} indeed has a non-zero value of $\partial P/\partial \Omega$ at $\Omega{=}0$, given by~\eqref{E:susceptibilities}.

Recently, Ref.~\cite{Nicolis:2011ey} argued for a relation between the angular momentum density and the Hall viscosity at $T=0$. It is worth pointing out that our holographic model of Section~\ref{S:holographic} has a non-zero value of $\MO=\partial P/\partial \Omega$ at $T\neq 0$, while the Hall viscosity $\tilde\eta$ vanishes identically.

\item{\bf The off-diagonal conductivity.}
    The total off-diagonal conductivity $\sigma_{xy}$, defined through $J^x = \sigma_{xy} E^y$, is given by
    \begin{equation}
    \label{E:xytotal}
        \sigma_{xy} = \tilde{\sigma} + \tilde{\chi}_E\,,
    \end{equation}
    where $\chiE$ is a thermodynamic response parameter and $\tilde{\sigma}$ is a transport coefficient.
    In the holographic example presented in Section~\ref{S:holographic}, $\tilde\sigma\neq 0$. We anticipate that $\tilde{\sigma}$ will contribute to an anomalous Hall effect \cite{nsomo} in parity-violating systems, and to an off-diagonal heat current, $Q^i = T^{0i} - \mu J^i$, at non-zero chemical potential. We point out that the total conductivity in \eqref{E:xytotal} contains contributions from both transport and bound currents. We plan to study these contributions in future work \cite{jkkmry}.

    The fluid-gravity correspondence allows an interesting interpretation of the off-diagonal conductivity in terms of properties of the black hole horizon. The fluid-gravity analysis of Section~\ref{S:holographic} can be extended to include a dilaton, $e(x^a)$, coupled to the gauge field via a $e^{-2}(x) F_{\mu\nu}F^{\mu\nu}$ term which replaces the canonical kinetic term. Such an extension will lead, at zero chemical potential, to the DC longitudinal conductivity $\sigma_{xx}=\sigma=1/e^2(r_H)$ along with the off-diagonal conductivity $\sigma_{xy}=\tilde{\sigma} + \tilde{\chi}_E=\theta(r_H)/(8\pi^2)$.\footnote{Note that a similar computation of the off-diagonal conductivity in a holographic system containing an axion was carried out in \cite{Iqbal:2008by}, obtaining a result for $\sigma_{xy}$ that differs from ours.}
    Since only the horizon value $r=r_H$ of the axion and dilaton enter the expressions for the conductivities this result may be suggestive of the membrane paradigm. In spite of this possible interpretation, one can show that at non-zero chemical potential the off-diagonal conductivity cannot be written in terms of horizon quantities.

\end{itemize}

\acknowledgments
\noindent
We thank D.~T.~Son for discussions which initiated this project.
We also thank F.~Benini, O.~Bergman, J.~Bhattacharya, S.~Bhattacharya, G.~Dunne, C.~P.~Herzog, G.~D.~Moore, K.~Landsteiner, S.~Minwalla, T.~Petkou, M.~Petropoulos, M.~Rangamani, D.~T.~Son and H.~Verlinde for helpful conversations.
KJ, PK, and AR are supported in part by NSERC, Canada.
MK was supported in part by the DFG (Deutsche Forschungsgemeinschaft) and is currently supported by the US Department of Energy under contract number DE-FGO2-96ER40956.
The research of RM is supported by the European Union grant FP7-REGPOT-2008-1-CreteHEPCosmo-228644.
AY is a Landau fellow, supported by the Taub foundation. AY is also supported in part by the Israeli Science Foundation (ISF) under grant number 495/11 and by the Bi-national Science Foundation (BSF) grant number 2014350. We thank a variety of institutions for their hospitality and support during the completion of this work, including the Aspen Center for Physics, the Galileo Galilei Institute in Florence, the Kavli Institute for Theoretical Physics, the Laboratoire de Physique Th\'eorique at the \'Ecole Normale Sup\'erieure, the Lorentz Center at the University of Leiden, and the Perimeter Institute for Theoretical Physics.

\bibliographystyle{JHEPx}
\bibliography{3dhydrobib}

\end{document}